\documentclass[twocol]{ametsocV6.1}

\usepackage{amsmath,amsfonts,amssymb,bm}
\usepackage{mathptmx}
\usepackage{newtxtext}
\usepackage{newtxmath}
\newcommand{\pder}[2][]{\frac{\partial#1}{\partial#2}}
\newcommand{\dder}[2][]{\frac{\text{d}#1}{\text{d}#2}}
\newcommand{\ddder}[2][]{\frac{\text{d}^2#1}{\text{d}#2^2}}

\newcommand{\dpder}[2][]{\frac{\partial^2#1}{\partial#2^2}}

\title{Upwards Tropospheric Influence on Tropical Stratospheric Upwelling \\
\textcolor{red}{Under peer review}}

\authors{Jonathan Lin\aff{a,c} \correspondingauthor{Jonathan Lin, jlin@ldeo.columbia.edu}
Kerry Emanuel, \aff{b}}

\affiliation{\aff{a} {Lamont-Doherty Earth Observatory, Columbia University, New York, New York} \\
\aff{b} {Lorenz Center, Department of Earth, Atmospheric, and Planetary Sciences,\\Massachusetts Institute of Technology, Cambridge, Massachusetts} \\
\aff{c} {Department of Earth and Atmospheric Sciences, Cornell University, Ithaca, New York}}

\abstract{The steady response of the stratosphere to a sea surface temperature (SST) forcing is considered in two separate theoretical models. It is first shown that anomalies in SST impose a geopotential anomaly at the tropopause. Solutions to the linearized quasi-geostrophic potential vorticity equations are then used to show that the vertical length scale of a tropopause geopotential anomaly is initially shallow, but significantly increased by diabatic heating from radiative relaxation. This process is a quasi-balanced response of the stratosphere to tropospheric forcing. A previously developed, coupled troposphere-stratosphere model is then introduced and modified. Solutions under steady, zonally-symmetric SST forcing in the linear $\beta$-plane model show that the upwards stratospheric penetration of the corresponding tropopause geopotential anomaly is controlled by two non-dimensional parameters, (1) a dynamical aspect ratio, and (2) a ratio between tropospheric and stratospheric drag. The meridional scale of the SST anomaly, radiative relaxation rate, and wave-drag all significantly modulate these non-dimensional parameters. Under Earth-like estimates of the non-dimensional parameters, the theoretical model predicts stratospheric temperature anomalies 2-3 larger in magnitude than that in the boundary layer, approximately in line with observational data. Using reanalysis data, the spatial variability of temperature anomalies in the troposphere is shown to have remarkable coherence with that of the lower-stratosphere, which further supports the existence of a quasi-balanced response of the stratosphere to SST forcing. These findings suggest that besides mechanical and thermal forcing, there is a third way the stratosphere can be forced -- through the tropopause.}

\begin{document}

\maketitle

%
%
%
\statement Upwards motion in the tropical stratosphere, the layer of atmosphere above where most weather occurs, is thought to be controlled by weather disturbances that propagate upwards and dissipate in the stratosphere. The strength of this upwards motion is important since it sets the global distribution of ozone. We formulate and use simple mathematical models to show the vertical motion in the stratosphere can also depend on the warming in the troposphere, the layer of atmosphere where humans live. We use the theory as an explanation for our observations of inverse correlations between the ocean temperature and the stratosphere temperature. These findings imply that stratospheric cooling may be tightly coupled to ocean warming.

%
%
%

%




\section{Introduction}


The Brewer-Dobson circulation (BDC) is a global-scale overturning circulation in the stratosphere, characterized by air that ascends into and within the tropical stratosphere, spreading poleward and eventually downwards in the extratropical winter-hemisphere. This stratospheric circulation can significantly impact tropospheric climate, most notably through its modulation of the distribution of stratospheric ozone, which absorbs harmful ultraviolet radiation from the sun \citep{dobson1956origin}. The widely accepted mechanism that explains the existence of the BDC is the principle of ``downward control” \citep{haynes1987evolution, haynes1991downward}. This principle states that for steady circulations, the upward mass flux across a specified vertical level is solely a function of the zonal momentum sources and sinks above that level; thus, processes in the middle and upper stratosphere can exert a “downward” influence on flow in the lower stratosphere and troposphere. In the stratosphere and mesosphere, it is primarily the dissipation of upward propagating Rossby and gravity waves that contributes zonal momentum \citep{seviour2012brewer}. The theoretical findings of \citet{haynes1991downward} have been well supported by numerical modeling evidence and withstood the test of time \citep[and references therein]{butchart2014brewer}. Thus, in the “downward control” paradigm, wave dissipation drives the circulation.

The BDC is typically separated into two branches: a slow and deep equator-to-pole overturning branch, and a faster shallow branch in the lower stratosphere extending to about 50$^\circ$ latitude \citep{plumb2002stratospheric, birner2011residual}. The deep branch is thought to be driven by planetary scale waves breaking in the middle and upper portions of the stratosphere, a process also known as the extratropical pump \citep{holton1995stratosphere}. The shallow branch is thought to be driven by sub-tropical wave-dissipation in the lower stratosphere \citep{plumb1999brewer, plumb2002stratospheric}. In this study, we focus primarily on the shallow branch circulation, and its connections to the tropical troposphere.

In our opinion, there are a few characteristics of the shallow branch circulation that remain unresolved. First, calculations of residual vertical velocities at 70-hPa indicate off-equator maxima in shallow branch upwelling in the summer-time hemisphere \citep{randel2008dynamical, seviour2012brewer}. Even though wave-drag can force circulations non-linearly and non-locally, wave-drag is at its annual maximum in the winter hemisphere, which is thus at odds with the observation of tropical upwelling maximizing in the summer-time hemisphere \citep{holton1995stratosphere, plumb1999brewer}. In fact, all of the experiments performed in \citet[hereafter, PE99]{plumb1999brewer} showed that as long as wave-drag maximizes in the winter hemisphere, upwelling maximizes in the winter hemisphere. Only when thermal forcing was included, did PE99 observe upwelling maximizes in the summer hemisphere. PE99 also found that the existence of a thermally-forced circulation in the stratosphere and the breakdown of downward control theory go together. This led PE99 to question the generality of downward control in the deep tropics, and whether or not thermally forced tropospheric circulations, such as the Hadley cell, could penetrate upwards into the stratosphere. Since the Hadley circulation is closely tied to the meridional gradient of sea surface temperature (SST) \citep{emanuel1995thermally}, the connection between tropospheric warming and lower stratospheric upwelling is one that perhaps deserves attention.


If tropospheric warming and stratospheric upwelling are connected, then of particular importance is the tropical tropopause layer (TTL), which serves as a boundary between the troposphere and stratosphere \citep{fueglistaler2009tropical}. Much research has focused on the temperature in the TTL region, since it has been linked with the concentration of water vapor in the stratosphere \citep{jensen2004transport, fueglistaler2005stratospheric, randel2006decreases, randel2019diagnosing}. In the tropical stratosphere, upwelling strength is strongly correlated with temperature, since a cold anomaly that slowly varies in time must be maintained by adiabatic cooling against the effect of radiative heating. Indeed, observational data suggests a strong link between the two \citep{randel2006decreases, kerr2006tropical}. Via downward-control arguments, wave-dissipation has been historically linked with tropopause temperature. For instance, an annual cycle in sub-tropical wave-dissipation of equatorward propagating extra-tropical waves has been suggested as responsible for the annual cycle in tropical tropopause temperature (which is much larger in amplitude than that of the tropical troposphere) \citep{yulaeva1994cause, holton1995stratosphere, randel2002time, taguchi2009wave, garny2011dynamically, kim2016spectrum}. Other studies have also attempted to understand how waves originating in the tropics (which can be excited by deep convection) can explain various aspects of the tropopause region, including the annual cycle in temperature \citep{boehm2003implications, norton2006tropical, randel2008dynamical, ryu2010effect, ortland2014residual, jucker2017untangling}. In this view, the strength of upwelling in the lower stratosphere is the primary control on temperature near the tropopause.


Changes to the tropopause temperature could theoretically induce changes in shallow branch upwelling, though a corresponding, self-consistent change in the momentum budget must also occur to balance the changes in the meridional circulation \citep{ming2016double}. In the tropics, many observational studies have found that, on a variety of space and time scales, strong cold anomalies often occur above regions of deep convection -- in essence, tropopause cooling is associated with tropospheric heating on the local and regional scale \citep{johnson1982thermodynamic, gettelman2002distribution, holloway2007convective, kim2012tropical, virts2014observations, kim2018convectively}. Some studies have argued that convection has a limited influence on the tropopause temperature since convection rarely penetrates the tropopause \citep{folkins1999barrier, gettelman2002distribution}. However, other studies have suggested that convection has a strong control on tropopause temperature, despite the rarity of tropopause-penetrating convection \citep{dessler2002effect, kuang2004convective}. Still, there is an oft-observed link between tropopause cooling and deep convection. In our view, there exists two theories that specifically explain this association. \citet{holloway2007convective} use a simple 2-D, linearized Boussinesq model to show that a convective ``cold-top" forms via ``hydrostatic adjustment" to convective heating. There is no dependence of the temperature anomaly on the horizontal scale in this theory. In contrast, it has also been argued that deep convection can excite a large-scale Kelvin wave response, which also has a vertically tilted signature of tropopause cooling \citep{kiladis2001aspects, randel2003thermal}. Most of these observational studies, however, focus on time scales much faster than that of the Brewer-Dobson circulation. But, there is also remarkable spatial correlation between tropospheric warming and stratospheric cooling trends on global warming time scales [see Fig. 1 of \citep{fu2006enhanced}].

If one persists with the assumption that the same mechanism responsible for local and regional scale anti-correlations between tropospheric warming and tropopause cooling can manifest itself at the global scale (which is not a given), then it is perhaps unsurprising that there also exists a tight coupling between tropospheric warming and the BDC shallow branch mass flux, at least when using SST to characterize the tropical troposphere. In general circulation models (GCMs) and re-analyses, there are strong correlations between tropical-mean SST and the BDC shallow branch mass flux, across a wide variety of time scales \citep{lin2015tropical, orbe2020giss, abalos2021brewer}. Fluctuations in tropical stratospheric upwelling have also been tied to ENSO (El Niño Southern Oscillation), one of the dominant sources of interannual tropical SST variability \citep{randel2009enso}. In fact, interannual variations in tropical mean SST explain ~40-50\% of the interannual variability of the 70-hPa vertical mass flux \citep{lin2015tropical, abalos2021brewer}. In addition, nearly 70\% of the CMIP6 model spread in the long-term trend of shallow branch mass flux is explained by the spread in tropical warming \citep{abalos2021brewer}. 

The tight coupling between tropical SST and BDC shallow branch upwelling on interannual to climate change time scales has been explained through changes to the wave-drag, in light of the downward-control paradigm: surface warming leads to upper tropospheric warming and modification of the sub-tropical jets, which can alter the upwards propagation and dissipation of mid-latitude waves in the sub-tropics \citep{garcia2008acceleration, calvo2010dynamical, shepherd2011robust, lin2015tropical}. While these theories (that are based on changes to zonal-mean wave-drag) can explain how SST and shallow branch mass flux are correlated, they were not constructed to also explain the oft-observed local-scale anti-correlation between SST and tropopause temperature.

\begin{figure}[t]
  \noindent\includegraphics[width=19pc]{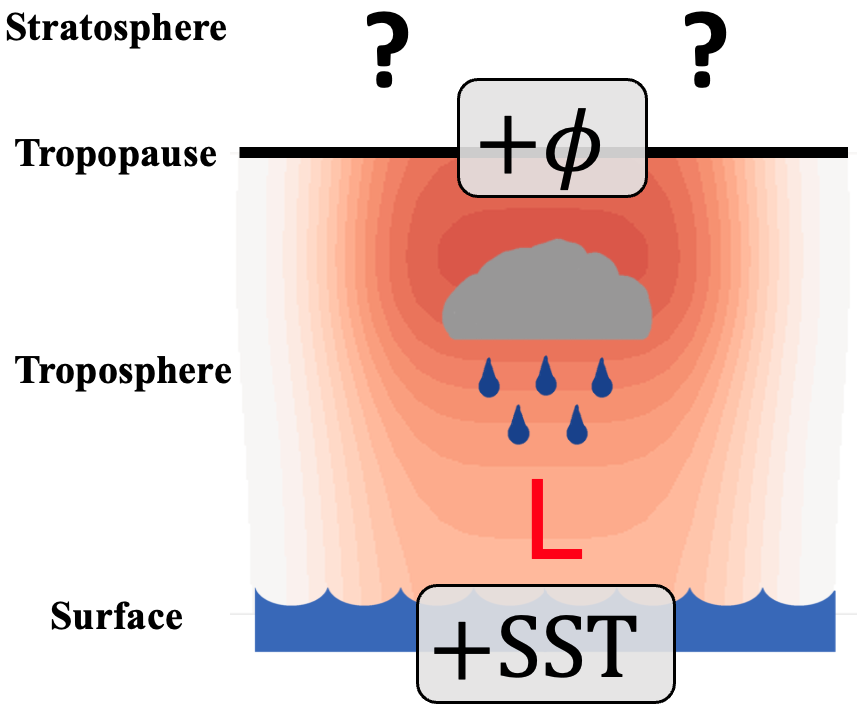}\\
  \caption{Schematic of a troposphere in radiative-convective-equilibrium, with an overlying stratosphere that is at rest. The troposphere is forced with a steady warm SST anomaly in the ocean. The troposphere warms (indicated by color shading) following a moist adiabat, the surface pressure falls, and the geopotential rises at the tropopause. How does the stratosphere respond to the an imposed tropopause geopotential anomaly?}
  \label{fig_schematic}
\end{figure}

In this study, we will put forth an alternative explanation for the anti-correlation between tropospheric and lower stratospheric temperature, and also attempt to understand the degree to which zonally-symmetric tropospheric heating can directly influence upwelling in the lower stratosphere. To start, consider the simplified atmospheric state shown in Figure \ref{fig_schematic}, which has a troposphere in radiative convective equilibrium, with an overlying stratosphere at rest. Suppose we impose a steady patch of positive SST anomaly in the ocean. The increased surface enthalpy flux warms the troposphere, following a moist adiabat. The surface pressure falls, and the geopotential at the tropopause rises. Since there cannot be a pressure discontinuity across the tropopause, the pressure must also rise in the lower stratosphere. How far up does it extend, and what is the steady response in the stratosphere?

Section \ref{sec_qgpv} tries to answer this conceptual question by introducing the concept of SST forcing at the tropopause and building a zonally asymmetric framework to understand the processes that control the upwards extent of tropopause anomalies. It is shown that there is a quasi-balanced response of the stratosphere to tropospheric forcing. Section \ref{sec_coupled} uses a steady, coupled troposphere-stratosphere system to show how zonally symmetric SST anomalies can influence tropical upwelling. Section \ref{sec_reanalyis} uses reanalysis data to argue for the real-world presence of the processes described in the proposed theory. Section \ref{sec_conclusion} concludes the study with a summary and discussion.

\section{Stratospheric Response to a Tropopause Anomaly \label{sec_qgpv}}
In this section, we introduce a simple conceptual model that will (1) illuminate how SST forcing can induce a tropopause geopotential anomaly, and (2) understand what parameters modulate the upwards extent of the tropopause anomaly into the stratosphere.

To understand how the stratosphere could be forced by the troposphere, we begin with tropospheric dynamics. In radiative-convective equilibrium, a valid approximation is that of strict convective quasi-equilibrium, where the saturation moist entropy, $s^*$, is constant with height ~\citep{emanuel1987air, emanuel1994large}. ~\citet{emanuel1987air} showed that linearized geopotential perturbations are directly connected to linearized $s^*$ perturbations (note here, for simplicity, we have ignored the small effect of water vapor on density):
\begin{equation}
    \pder[\phi^\prime]{p} = -\bigg( \pder[T]{p} \bigg)_{s^*} s^{*\prime} \label{eq_dphi_dp}
\end{equation}
where prime superscripts indicate perturbation quantities. Since $s^*$ is constant with height, Eq. \ref{eq_dphi_dp} can be directly integrated in pressure to yield (as also shown in \citet{lin2022effect}):
\begin{equation}
    \phi^\prime(p) = \phi^\prime_b + s^{*\prime} (\overline{T}_b - \overline{T}(p)) \label{eq_phi_qe}
\end{equation}
where $\phi^\prime_b$ is the perturbation boundary layer geopotential, $\overline{T}$ is the basic state temperature, and $\overline{T}_b$ is the basic state boundary layer temperature. We non-dimensionalize according to:
\begin{equation}
    \phi \rightarrow H^2 N^2 \phi \qquad s^* \rightarrow \frac{H^2 N^2}{\overline{T}_b - [\overline{T}]} s^* \label{eq_nd_sphi}
\end{equation} 
where $H$ is the scale height , $N^2$ is the buoyancy frequency, and $[\overline{T}]$ is the basic state vertically-averaged temperature. Dropping primes for perturbation quantities and non-dimensionalizing yields:
\begin{equation}
    \phi(p) = \phi_b + \big(1 - V_1(p) \big) s^* \label{eq_phi_p}
\end{equation}
where $V_1$ is the non-dimensional first baroclinic mode \citep{lin2022effect}: 
\begin{equation}
    V_1(p) = \frac{\overline{T}(p) - [\overline{T}]}{T_{b} - [\overline{T}]} \label{eq_baroclinic_mode}
\end{equation}
Eq. \ref{eq_baroclinic_mode} shows that the first baroclinic mode is positive near the surface, transitions to zero in the mid-troposphere, and is negative at the tropopause. Evaluating Eq. \ref{eq_phi_p} at the tropopause yields:
\begin{equation}
    \phi(\hat{p}_t) = \phi_0 - V_1(\hat{p}_t) s^* \label{eq_phi_qe_trop}
\end{equation}
where $\hat{p}_t$ is the non-dimensional tropopause pressure, and $\phi_0 = \phi_b + s^*$ is the barotropic geopotential. Note, the barotropic geopotential is constant with height. The total geopotential is the linear sum of the contributions of the tropospheric barotropic and baroclinic geopotential.

Since the tropopause is colder than the mean troposphere temperature, $V_1(\hat{p}_t)$ is negative, such that for positive SST anomalies ($s^{*\prime} > 0$), the tropopause geopotential anomaly will also be positive, provided the barotropic geopotential is not less than $V_1(\hat{p}_t) s^*$. In the real atmosphere,  baroclinic perturbations are typically around an order of magnitude larger than barotropic ones \citep{lin2022effect}, such that for the sake of simplicity, we proceed with the approximation that $\phi_0$ is small in relation to the baroclinic term. We will relax this assumption in the next section. Then, in this simple conceptual framework, we have a warm patch of ocean that imposes a steady positive geopotential anomaly at the tropopause.

Next, we will consider what happens to the stratosphere subject to a steady tropopause forcing (i.e. a steady lower boundary condition). The response of the stratosphere to external forcing has been well-studied using theoretical models [see \citet{garcia1987mean, haynes1991downward, plumb1999brewer}, among many others]. However, the external forcing is typically presented in terms of being mechanical (wave-driven) or thermal in origin. We instead impose a tropopause forcing via the SST anomaly, and use the well-known quasi-geostrophic potential vorticity equations (QGPV), linearized about a resting basic state on an f-plane:
\begin{align}
    q^\prime(x, y, z) &= \frac{1}{f_0} \nabla_H^2 \phi^\prime + \frac{f_0}{N^2} \dpder[\phi^\prime]{z} - \frac{f_0}{H N^2} \pder[\phi^\prime]{z}
\end{align}
where $q$ is the potential vorticity (PV), $f_0$ is the Coriolis parameter, $N$ is the buoyancy frequency, $\phi$ is the geopotential. Here, we are considering perturbations large enough in scale for the quasi-geostrophic approximation to apply. Dropping primes for perturbation quantities, assuming wave-like solutions in the zonal and meridional [$\exp(ikx + ily)$], and non-dimensionalizing by:
\begin{align}
\begin{split}
    &x \rightarrow L x \qquad y \rightarrow L y \qquad z \rightarrow H z  \\
    &\phi \rightarrow H^2 N^2 \phi \qquad q \rightarrow f_0 q \qquad t \rightarrow t / f_0 \label{eq_nd}
\end{split}    
\end{align}
where $L = N H / f$ is the Rossby radius of deformation, we obtain
\begin{equation}
    \Big(\dpder[]{z} - \pder[]{z} -(k^2 + l^2) \Big) \phi = q(z) \label{eq_qgpv_linear}
\end{equation}
These equations can be found in most standard textbooks, e.g. section 5.4 of \citet{vallis2017atmospheric}. Here, we emphasize the boundary conditions:
\begin{align}
    \phi(z = 0) &= \phi_T \\
    \pder[\phi]{z}(z = \infty) &= 0
\end{align}
where the bottom boundary condition enforces continuity of pressure across the tropopause, given the aforementioned tropopause geopotential anomaly that is imposed by an SST anomaly. The upper boundary condition requires the temperature anomaly (or vertical velocity anomaly) be zero. Though $\phi_T$ is imposed by the troposphere, via Eq. \ref{eq_phi_qe_trop}, in reality, barotropic motions are coupled to the stratosphere. Thus, we can only assume the geopotential as a steady lower boundary condition, and solve for the stratosphere in isolation, since we ignored the barotropic geopotential. As shall be illuminated in the next section, the barotropic mode should really be coupled to the stratospheric circulation. 

We proceed by considering the stratospheric response to a geopotential anomaly at the tropopause, with zero perturbation PV throughout the rest of the stratosphere. Since imposing a geopotential anomaly at the tropopause has no direct effect on stratospheric PV, it can be considered as the fast stratospheric response to a tropopause geopotential anomaly. In this textbook case, the solution is straightforward:
\begin{equation}
    \phi(z) = \exp \big( m_- z \big)
\end{equation}
where
\begin{equation}
    m_- = \frac{1 - \sqrt{1 + 4(k^2 + l^2)}}{2}
\end{equation}
which shows that the geopotential anomaly decays in the vertical with a scale inversely proportional to the horizontal scale of the anomaly. On re-dimensionalization, the Rossby penetration depth,
\begin{equation}
    R_d = \frac{f_0 L}{N}
\end{equation}
where $L$ is the horizontal scale, is the operative vertical scale of the geopotential. Tropopause anomalies with large horizontal scales will extend deeper into the stratosphere than smaller ones.

The temperature anomaly, scaling with $\pder[\phi]{z}$, will also decay exponentially with height according to $R_d$. But how large can the temperature anomalies get? Thermal wind balance dictates that
\begin{equation}
    g \pder[\ln T]{y} = -f \pder[u]{z}
\end{equation}
If we take $\partial z$ to scale as the Rossby penetration depth, then we obtain:
\begin{equation}
    \ln T \approx \frac{N u}{g}
\end{equation}
Note that $f$ drops out, which indicates that the temperature in the stratosphere does not directly depend on $f$. It rather depends on the magnitude of the tropopause anomaly, as well as the stratospheric stratification. For the case of zero perturbation PV in the stratosphere, the temperature anomaly is just the geopotential anomaly multiplied by $m_-$, which is inversely proportional to the horizontal scale of the tropopause PV anomaly. Therefore, the magnitude of the tropopause temperature perturbations can be large for small horizontal scale anomalies, though these will be confined to a rather shallow vertical layer near the equator (and may also not obey the quasi-geostrophic approximation).

Next, it is instructive to consider how the stratosphere responds to the temperature anomalies. As alluded to earlier, temperature anomalies disturb the radiative equilibrium of the stratosphere. This must be associated with radiative heating anomalies. In this case, PV is no longer conserved. The response of the stratosphere can be modeled as:
\begin{equation}
    \pder[q]{t} = \frac{f_0}{N^2} \pder[\dot{Q}]{z} \label{eq_dqdt}
\end{equation}
where $\dot{Q}$ is the heating rate (thermal forcing), and is parameterized to be a simple Newtonian radiative relaxation:
\begin{equation}
    \dot{Q} = - \alpha_{\text{r}} \pder[\phi]{z} \label{eq_heating}
\end{equation}
$\alpha_{\text{r}} > 0$ is the inverse time scale of the Newtonian radiative relaxation. \citet{hitchcock2010approximation} found that linear radiative relaxation can explain around 80\% of the variance in longwave heating rates in a climate model, though this is less accurate in the lower stratosphere, and dependent on the relaxation rate having a height-dependence. Non-dimensionalizing using Eq. \ref{eq_nd}, we obtain:
\begin{equation}
    \pder[q]{t} = -\gamma \dpder[\phi]{z} \label{eq_dqdt_nd}
\end{equation}
where $\gamma = \alpha_{\text{rad}} / f_0$.

The effect of radiative damping on stratospheric circulations has been thoroughly explored in a number of early theoretical studies \citep{garcia1987mean, haynes1991downward, haynes1993effect}. In particular, the seminal work of \citet{haynes1991downward} showed that in zonally symmetric, radiatively damped, time-dependent systems whereby a steady mechanical forcing is instantaneously applied, there is an adjustment to a barotropic state (in $u$) above the level of forcing. Our set up is similar to the model outlined in section 3 of \citet{haynes1991downward}, except here the steady forcing is restricted to the tropopause geopotential -- the forcing is neither wave-driven nor thermal in origin.

\begin{figure}[th]
  \noindent\includegraphics[width=19pc]{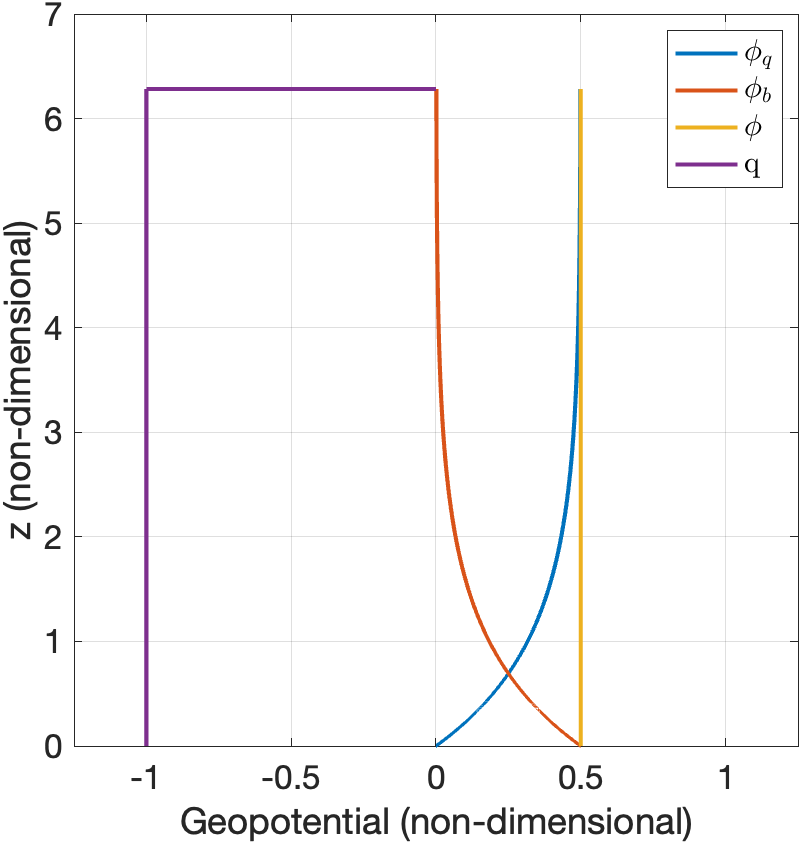}\\
  \caption{The geopotential associated with (red) a boundary PV anomaly of $q = -1$ ($\phi_b$), (blue) a constant PV anomaly of $q = -1$ in the interior ($\phi_q$), and (yellow) the sum of the two ($\phi = \phi_q + \phi_b)$. The corresponding total PV is shown in purple. Here we assume $k_m = 2$, and $z_{\text{top}} = 1 + 2 \pi$.}
  \label{fig_barotropic_solution}
\end{figure}

To solve for the geopotential, the Green's function (see the Appendix) is convoluted with the source term under the lower boundary condition:
\begin{equation}
    q_T = -k_m \phi_T
\end{equation}
where $k_m = k^2 + l^2$ is the total wavenumber. This can be calculated numerically (see the Appendix for more details). Figure \ref{fig_barotropic_solution} shows the stratospheric geopotential solutions that describe the initial and final states after imposing a tropopause geopotential anomaly. The initial geopotential distribution from the steady geopotential anomaly is shown as $\phi_b$, and is just the zero interior perturbation PV solution mentioned earlier in the text, where the response decays exponentially with height. The geopotential distribution associated with the generation of anomalous PV through diabatic heating by radiative relaxation is shown in $\phi_q$, while the total geopotential is shown as $\phi = \phi_q + \phi_b$. The total geopotential is constant with height (barotropic) above the level of forcing, as found by \citet{haynes1991downward}.

A simple physical picture is painted with this conceptual model that can provide an rather straightforward answer to the schematic shown in Figure \ref{fig_schematic}. If the troposphere is forced with a steady positive SST anomaly, a positive geopotential anomaly forms at the tropopause. A positive tropopause geopotential anomaly is initially accompanied with a cold anomaly in the stratosphere, which is associated with radiative heating and rising motion. If this process is allowed to proceed towards a steady state back to radiative equilibrium, the geopotential and PV must eventually become constant with height (i.e. barotropic), as implied by Eq. \ref{eq_heating}. In this way, the troposphere can force the stratosphere, at least on the steady time scales considered here. This also shows that the geopotential does not have to go to zero at the upper boundary. The only requirement is that the energy density goes to zero. Thus, the assumption of the geopotential going to zero at the upper boundary in \citet{holloway2007convective} seems arbitrary.

How long does it take to reach the barotropic state? \citet{haynes1991downward} showed that in the zonally symmetric case, the adjustment towards a barotropic state above the level of forcing occurs with an upward propagation speed of $w_\alpha = \alpha_{\text{rad}} R_d^2 / H_s$. In the tropics, $w_\alpha$ is small, owing to the smallness of both $\alpha_{\text{rad}}$ and $R_d$. For an anomaly of horizontal scale around 5000~km at a latitude of $10^\circ$, and a radiative relaxation time scale of $\alpha_{\text{rad}} = 20$ days$^{-1}$, $w_\alpha \approx O(10^{-1})$ mm s$^{-1}$. This corresponds to an upward propagation of only a few km per year. It is also possible to numerically calculate the amount of time it takes for the system to reach its final barotropic state, by time-stepping Eq. \ref{eq_dqdt_nd} forwards in time while holding the lower-boundary PV fixed. For a stratosphere with a depth of around 32-km ($z_{\text{top}} = 4$ for a scale height of $H_s = 8$ km), assuming $\gamma = 0.02$ and a Coriolis parameter akin to that at 10$^\circ$ latitude, it takes around 3 years for the system to become barotropic.

This long relaxation time makes it unlikely that the barotropic state is ever reached in the real stratosphere, since unsteady processes can disrupt the simple state assumed in this model. For instance, it is unlikely that a tropopause geopotential anomaly would remain steady on the order of years. Furthermore, since the $\beta$-effect is not included in this simple framework, we also ignore the possibility of the excitation of large-scale waves (and their corresponding effects) as a part of the response to the tropopause geopotential anomaly.

\begin{figure*}[th]
  \noindent\includegraphics[width=39pc]{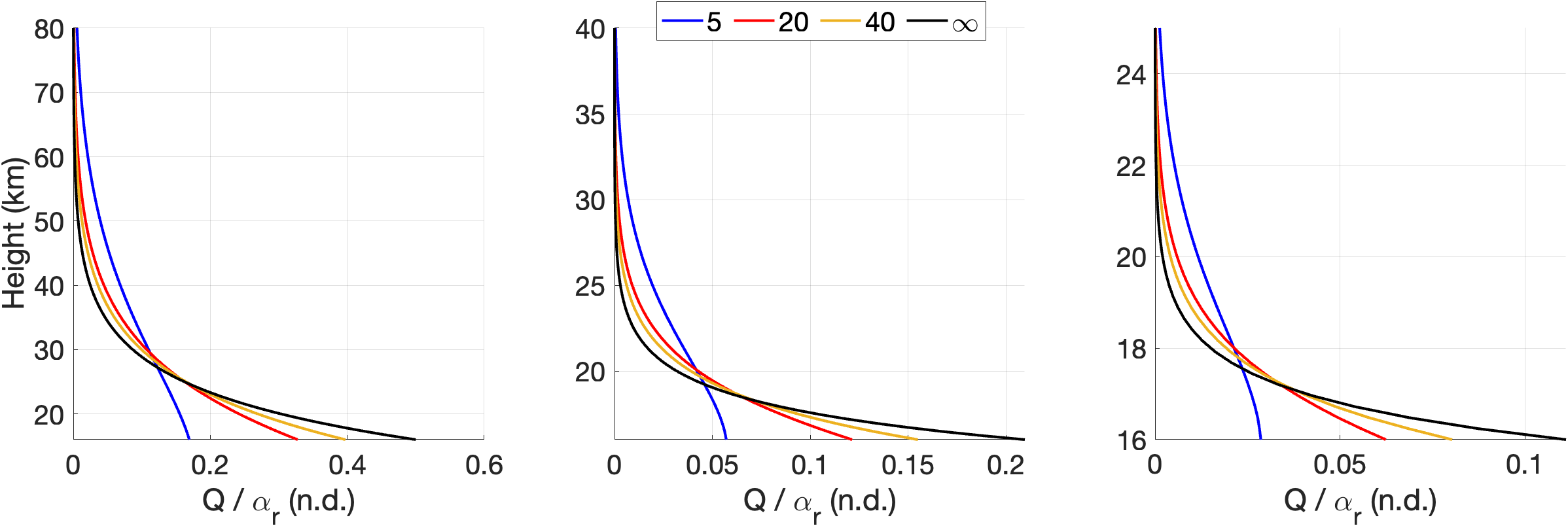}\\
  \caption{(Left) The diabatic heating profile ($Q / \alpha_r$) with height in the stratosphere after 30 days of integration, subject to a steady tropopause boundary forcing with a horizontal scale of around $28000$-km, and a (blue) 5-day, (red) 20-day, (yellow) 40-day. The vertical derivative of the geopotential for the zero-PV stratospheric response to a tropopause forcing (infinite radiative relaxation time scale) is shown in black. (Middle) and (Bottom) are the same as top but for a horizontal scale of around $9500$-km, and $4500$-km, respectively. We assume a scale height of $8$ km, and a tropopause height of 16 km to convert to dimensional height.}
  \label{fig_diabatic_30d}
\end{figure*}

In light of this, the intermediate states between the fast stratospheric response [$\phi_b$ in Figure \ref{fig_barotropic_solution}] in which the anomaly decays exponentially with height, and the barotropic steady-state response in which the boundary anomaly is communicated throughout the depth of the stratosphere [$\phi$ in Figure \ref{fig_barotropic_solution}], could be important. For practical purposes, the geopotential anomaly is not as important as the associated radiative heating, which is potentially important for tracer transport into the stratosphere. Figure \ref{fig_diabatic_30d} shows the non-dimensional diabatic heating profiles with height after 30 days of integration, for a stratosphere subject to an imposed tropopause geopotential anomnaly that is associated with a unitary non-dimensional anticyclonic PV, under varying magnitudes of stratospheric radiative relaxation rates. The diabatic heating profiles are normalized by the radiative relaxation rate. For comparison purposes, we show the temperature anomaly associated with the (time-independent) zero perturbation PV geopotential solution (i.e. an infinite radiative-relaxation time scale), even though there is no associated diabatic heating, by definition. Figure \ref{fig_diabatic_30d} shows that after 30-days, there is non-trivial lifting (in height) of the diabatic heating anomaly over time. The stronger the strength of radiative relaxation, the faster the diabatic heating anomaly is communicated into the stratosphere.

These calculations show that tropospheric heating imposes a positive tropopause geopotential anomaly, which elicits a quasi-balanced response in the stratosphere. The fast stratospheric response is simply an anomaly that decays in the vertical according to the Rossby penetration depth. On slower time scales, radiative relaxation induces an upward migration of the anomaly. The radiative relaxation rate, the horizontal scale of the anomaly, and the Coriolis parameter all determine the upward migration rate, as shown in \citet{haynes1991downward}. Thus, the ensuing, time-dependent temperature response in the stratosphere is also tied to these parameters. In the next section, we will elaborate on the ideas put forth in this conceptual model in a zonally-symmetric framework, and analyze, in detail, the sensitivity of the stratospheric response to tropospheric forcing, with regards to these parameters.

\section{Troposphere-Stratosphere Response to SST \label{sec_coupled}}
In the previous section, we used a simple QGPV framework to understand how a SST anomaly can impose a tropopause geopotential anomaly and therefore elicit a quasi-balanced response in the stratosphere. However, we used the tropopause as a lower boundary condition for the stratosphere when in reality, the tropopause and stratosphere are coupled. In this section, we develop a simple, zonally-symmetric, coupled troposphere-stratosphere model, and explore how radiation and wave-drag can modulate the response of the stratosphere to SST forcing.

\subsection{Model Formulation}
\citet{lin2022effect} formulated a linear, coupled troposphere-stratosphere model, but in the context of unsteady equatorial waves. In that linear system, a convecting, quasi-equilibrium troposphere was coupled to a dry and passive stratosphere. We use the same non-dimensional system derived in \citet{lin2022effect}, except we only consider steady, zonally symmetric circulations. The tropospheric system is governed by:
\begin{align}
     y v_0 - F (u_0 + u_1) &= 0 \label{eq_steady_uBT} \\
    -\pder[\phi_0]{y} - y u_0 &= 0 \label{eq_steady_vBT} \\
    y v_1  - F (u_0 + u_1) - D_t u_1 &= 0 \label{eq_steady_uBC} \\
    y u_1 &= \dder[s^*]{y} \label{eq_steady_vBC} \\
    \pder[v_0]{y} + \pder[v_1]{y} + \pder[\omega]{y} &= 0 \label{eq_cont}
\end{align}
where $u_0$ and $v_0$ are the barotropic zonal and meridional winds (constant with height), $u_1$ and $v_1$ are the baroclinic zonal and meridional winds, $\phi_0$ is the barotropic geopotential, $s^*$ is the saturation moist entropy (that is assumed to be vertically constant, as in a quasi-equilibrium troposphere), $D_t$ is a non-dimensional Rayleigh damping coefficient, and
\begin{equation}
    F = \frac{a C_d |\overline{\textbf{V}}|}{\beta L_y^2 h_b}
\end{equation}
is a non-dimensional surface friction coefficient (derived in \citet{lin2022effect}), where $C_d$ is the drag coefficient, $h_b$ is the boundary layer depth, $L_y$ is the meridional length scale, $\beta$ is the meridional gradient of the Coriolis force, $a$ is the radius of the Earth, and $\overline{\textbf{V}}$ is the basic state surface wind speed magnitude. The vertical structure of the baroclinic variables are determined by $V_1$ (Eq. \ref{eq_baroclinic_mode}). Note that while there are equations for the tropospheric thermodynamics in \citet{lin2022effect}, they are omitted here. Since $s^*$ is taken to be specified, representative of a SST forcing, there are 6 unknown variables, ($u_0$, $u_1$, $v_0$, $v_1$, $\omega$, $\phi_0$) and 5 equations. The system will be completed with a formulation of boundary conditions that will couple the troposphere system to a stratosphere (and provide the last equation).

In the ensuing text, terms with an overlying hat are dimensional. $\hat{D}_t$, the (dimensional) inverse time scale of the Rayleigh damping coefficient is:
\begin{equation}
    \hat{D}_t \rightarrow \frac{\beta L_y^2}{a} D_t
\end{equation}
In Eq. \ref{eq_steady_uBC}, $D_t u_1$ acts as a relaxational wave drag on the zonal flow. It does not act on the coupling between the troposphere and stratosphere, and is only used to diagnose $v_1$ (which by definition, has a value of zero at the tropopause). Thus, $D_t$ modulates the baroclinic vertical velocity profile in the zonally symmetric meridional overturning circulation. 

As formulated, the tropospheric system represents an atmosphere in which temperature anomalies in the vertical are restricted to follow the moist adiabat. The associated baroclinic mode, which is forced through surface enthalpy fluxes ($s^*$), can then excite the barotropic mode through surface friction \citep{lin2022effect}. The barotropic mode then excites the stratosphere. However, the stratospheric circulation becomes uncoupled with the tropospheric circulation when $F = 0$ -- in this case, the tropospheric solution simply obeys Eqs. \ref{eq_steady_uBC}-\ref{eq_cont}, and the barotropic mode (as well as the stratospheric state to tropospheric forcing) becomes ill-defined. This may imply that friction has an outsized influence on stratospheric circulations. This may not be true in reality, since the barotropic mode can also be coupled to the baroclinic mode through non-linearity and vertical wind shear. Both of these processes are not represented in this work.

The stratosphere is formulated in log-pressure coordinates and assumed to be in hydrostatic balance [see Chapter 3 of~\citet{andrews1987middle}]. The steady, linear, zonally symmetric, non-dimensional equations of the stratosphere are also derived from the system used in \citet{lin2022effect}, and summarized below:
\begin{align}
    y v_s - D_s u_s &= 0 \label{eq_strat_U_nd} \\
    -\pder[\phi_s]{y} - y u_s \label{eq_strat_V_nd} &= 0 \\
    \pder[v_s]{y} + \frac{1}{\rho_s} \pder[(\rho_s w_s)]{z^*} &= 0 \label{eq_strat_cont_nd} \\
    w_s S = -\alpha_{\text{rad}} \pder[\phi_s]{z} \label{eq_strat_T_nd} \\
    \rho_s = \exp \Big( \frac{H}{H_{s,s}}(1 - z^*) \Big) \label{eq_strat_rho}
\end{align}
where subscripts denote quantities in the stratosphere, $w_s$ is the log-pressure vertical velocity, $S$ is a non-dimensional stratospheric stratification, $\rho_s$ is the basic state density, $H$ is the dimensional tropopause height, $H_{s,s}$ is the dimensional scale height in the stratosphere, the log-pressure vertical coordinate ${z^* \equiv - H \ln ( p / p_t ) + 1}$ is defined such that $z^* = 1$ is the bottom boundary, or the tropopause, and $\alpha_{\text{rad}}$ is the non-dimensional radiative damping time scale in the stratosphere:
\begin{equation}
    \hat{\alpha}_{\text{rad}} \rightarrow \frac{\beta L_y^2}{a} \alpha_{\text{rad}}
\end{equation}
Relaxational wave drag, $D_s u_s$, is included only in the zonal momentum equations, as similarly used by \citet{plumb1999brewer}. It is not necessary that $D_s = D_t$, though discontinuities in the meridional velocity at the tropopause will occur if $D_s \neq D_t$. Furthermore, while this form of wave drag is simplistic, it is a rather poor representation of the response of the circulation to external forces \citep{ming2016response}. 

Finally, $S$ plays an important role in the behavior of this model, and is:
\begin{equation}
     S = \frac{N^2 H^2}{\beta^2 L_y^4}
\end{equation}
where $N$ is the buoyancy frequency. Note, there is no explicitly imposed thermal or mechanical forcing in the stratosphere. Thus, we consider a stratosphere entirely forced from the troposphere.

\subsection{Stratospheric response to tropopause forcing}
In the case of an isolated stratosphere subject to a tropopause forcing, the stratospheric equations can be reduced to a single differential equation for the geopotential:
\begin{equation}
     \dpder[\phi]{z} - \frac{H}{H_{s,s}} \pder[\phi]{z} + \frac{\xi}{y^2} \bigg[ \dpder[\phi]{y} - \frac{2}{y} \pder[\phi] {y} \bigg] = 0 \label{eq_phi_strat}
\end{equation}
where
\begin{equation}
    \xi = \frac{D_s S}{\alpha_{\text{rad}}} = \frac{\hat{D}_s}{\hat{\alpha}_{\text{rad}}} \frac{N^2 H^2}{\beta^2 L_y^4}
\end{equation}
is a non-dimensional term that depends on the ratio between the time scale of wave-drag to that of radiation. This quantity is equivalent to a ``dynamical aspect ratio" that describes the ratio of the vertical to horizontal scale of the circulation response to an imposed forcing \citep{garcia1987mean, plumb1999brewer, haynes2005stratospheric, ming2016response}. As detailed in \citet{ming2016response}, who incorporated an additional external heating in the stratosphere, when the aspect ratio is large ($\xi >> 1$), the external heating is narrow and primarily balanced by upwelling, and when the aspect ratio is small ($\xi << 1$), the external heating is broad and primarily balanced by Newtonian cooling. In this study, the interpretation of $\xi$ does not have exactly the same meaning, since we do not impose a temperature-independent external heating to the system (which in the real world would arise from absorption of radiation by ozone) -- our simple system is instead forced via the tropopause geopotential, and upwelling always balances Newtonian cooling. Here, $\xi$ better describes the geopotential response with height. As we shall see later, when the radiative time scale is much faster than the wave-drag time scale ($\xi << 1$), the meridional derivative terms are small and the system will become nearly barotropic in the vertical. On the other hand, when the wave-drag time scale is much faster than the radiative time scale ($\xi >> 1$), the stratospheric signature of the tropopause anomaly is muted. Note the presence of $L_y$, which indicates the importance of the horizontal scale of the anomaly.

\begin{figure*}
  \noindent\includegraphics[width=39pc]{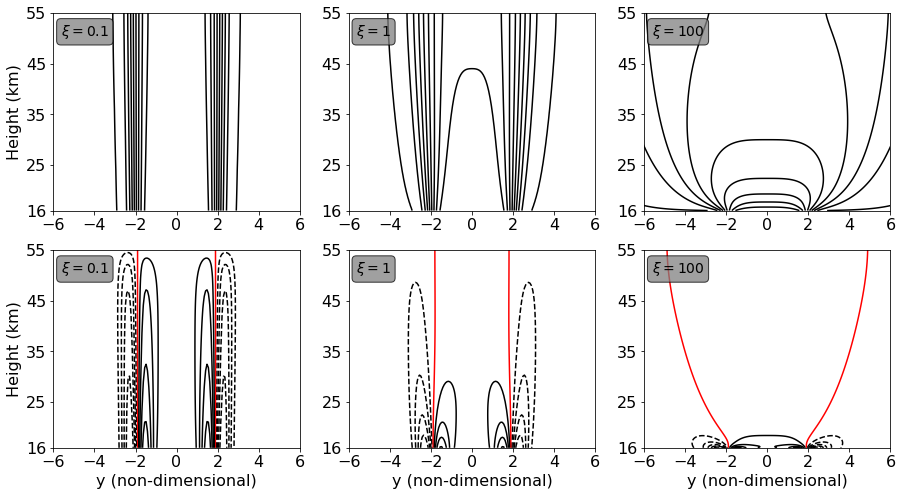}\\
  \caption{(Top-row) The zonally symmetric geopotential response to an imposed tropopause geopotential anomaly, as shown in Eq. \ref{eq_phi_overturning}, for varying values of $\xi$. (Bottom-row) Same as the top-row except for the zonally symmetric vertical velocity response. The red-line is the zero vertical velocity isoline. Tropopause height is $16$-km, and stratospheric scale height is $8$-km.}
  \label{fig_strat_zonally_symmetric_response}
\end{figure*}

Eq. \ref{eq_phi_strat} can be solved numerically, discretizing the grid in the meridional and vertical directions. The stratospheric geopotential is also subject to a zero temperature anomaly at the top of the domain, or equivalently, zero derivative of the geopotential. The geopotential anomaly is enforced to be zero on the northern and southern borders. For illustrative purposes, we first solve the equations under a fixed lower boundary condition:
\begin{equation}
    \phi(z^* = 1) = \phi_T
\end{equation}
where
\begin{equation}
    \phi_T = \int_y y \exp \big(-4 (y - 2)^2 \big) - y \exp \big(-4 (y + 2)^2 \big) \label{eq_phi_overturning}
\end{equation}
This represents a flat positive geopotential anomaly in the tropics (tropical heating) that decays to zero in the subtropics. As will become clear later when the solutions are coupled to the troposphere, this geopotential structure is associated with sub-tropical jets at $y = \pm 2$.

Figure \ref{fig_strat_zonally_symmetric_response} shows the stratospheric response to a tropopause geopotential anomaly, under varying values of $\xi$. Here, the numerical calculations confirm the mathematical analysis. Indeed, for $\xi = 0.1$ (i.e. when wave-drag is very weak), radiation acts to create a nearly barotropic stratosphere, in which motion is confined to constant angular momentum surfaces. The vertical structure of the vertical velocity in this case is qualitatively similar to the thermally forced vertical mode calculated in PE99 [see their Fig. 11]. When the time scale of wave-drag is faster than radiation ($\xi = 100$), the vertical penetration of the tropopause geopotential anomaly is significantly muted. In fact, the vertical velocity anomalies only extend on the order of a few km into the stratosphere. In this sense, the relaxational wave-drag acts to both mute the vertical scale of the tropopause geopotential anomaly, and sustain a meridional overturning circulation. 

As elaborated on earlier, there is much existing theoretical work that shows the response of the stratosphere to an external forcing is dependent on the strength of wave-drag, the strength of radiative relaxation, and the aspect ratio of the tropopause anomaly \citep{garcia1987mean, haynes1991downward, ming2016response}. This work is mathematically similar to and agrees with the aforementioned studies. Unlike the others, this work emphasizes the role of tropopause forcing on the stratosphere, and introduces the idea that there is a quasi-balanced response in the stratosphere to tropopause forcing, via tropospheric heating.

\subsection{Tropospheric forcing of stratospheric upwelling}
Next, we couple the stratospheric equations to the zonally symmetric tropospheric equations, to show how tropospheric thermal forcing can influence stratospheric upwelling. In order to couple the troposphere and stratosphere, we use classical matching conditions: (1) continuity of pressure (geopotential) and (2) vertical velocity at the tropopause:
\begin{align}
    \phi_s(z^* = 1) &= \phi_T \label{eq_phi_match} \\
    B \omega(\hat{p}_T) &= -w_s(z^* = 1) \label{eq_w_match}
\end{align}
where $B - \frac{H_{s,t}}{H} \frac{p_s - p_t}{p_t}$ is a scaling coefficient between pressure velocity and vertical velocity \citep{lin2022effect}. Here, $p_s$ is the surface pressure, $p_t$ is the tropopause pressure, and $H_{s,t}$ is the scale height of the troposphere. Solving for $v_0$ using Eqs. \ref{eq_cont}, \ref{eq_phi_match}, \ref{eq_w_match}, and assuming zero flow at the boundaries, yields:
\begin{equation}
    v_0 = \frac{\alpha_{\text{rad}}}{S B} \int_y \pder[\phi_s]{z} \bigg|_{z^* = 1} \: dy \label{eq_v0_strat}
\end{equation}
Here we see that under a rigid lid condition, where $S \rightarrow \infty$, $v_0 = 0$. In addition, $B$ is proportional to the troposphere scale height, which itself is inversely proportional to the dry stratification of the troposphere. Hence, $S B$ can also be thought of as a scaled ratio of the troposphere buoyancy frequency to the stratosphere buoyancy frequency. The strength of radiative relaxation also appears in the numerator. This is because the magnitude of the tropospheric barotropic mode is determined, in part, by stratospheric dynamics.

Eqs. \ref{eq_steady_uBT} and \ref{eq_steady_vBC} are used to solve for $u_0$ in terms of the stratosphere and the external forcing:
\begin{equation}
    u_0 = y \frac{1}{\xi \gamma} \int_y \pder[\phi_s]{z^*} \bigg|_{z^* = 1} \: dy - \frac{1}{y} \dder[s^*]{y} \label{eq_u0_coupled}
\end{equation}
where
\begin{equation}
    \gamma = \frac{F B}{D_S}
\end{equation}
is an additional non-dimensional parameter that qualitatively represents the ratio between stratospheric and tropospheric drag (there is tropospheric wave drag, but it does not act on the barotropic mode, only on the baroclinic mode). $\gamma$ is not entirely independent from $\xi$, since $D_s$ appears in both. Again, under the rigid lid condition, $\xi \rightarrow \infty$, such that the barotropic zonal wind becomes only a function of the tropospheric forcing. Note again that when $F = 0$, the barotropic mode becomes ill-defined, since it is no longer coupled to the baroclinic mode.


In order for the continuity of pressure to be satisfied, the geopotential at the lower boundary of the stratosphere must satisfy Eqs. \ref{eq_phi_qe_trop} and \ref{eq_phi_match}. Combining Eqs. \ref{eq_phi_qe_trop}, \ref{eq_steady_vBT}, \ref{eq_phi_match}, and \ref{eq_u0_coupled} yields:
\begin{multline}
    \pder[\phi_s]{y} \bigg|_{z^* = 1} \: - y^2 \frac{1}{\xi \gamma} \int_y \Big( \pder[\phi_s]{z} \Big) \bigg|_{z^* = 1} \: dy = \\ (1 - V_1 (\hat{p}_t)) \dder[s^*]{y} \label{eq_coupled_phi_s}
\end{multline}
which is an equation for the boundary geopotential entirely in terms of the external forcing, $s^*$. The Rayleigh damping coefficient for stratospheric wave-drag does not appear in the boundary condition, since
\begin{equation}
    \frac{1}{\xi \gamma} = \frac{\alpha_{\text{rad}}}{D_s S} \frac{D_s}{FB} = \frac{\alpha_{\text{rad}}}{S F B}
\end{equation}
When $\xi \gamma$ is large, the boundary condition simply reduces to Eq. \ref{eq_phi_qe_trop}, with $\phi_b = 0$. When $\xi \gamma$ is small, $s^*$ becomes a multiple of a double integral in $y$ of the vertical derivative of the stratospheric geopotential at the tropopause.

Incorporating Eq. \ref{eq_coupled_phi_s} as the lower boundary condition is numerically tricky given the meridional integral, since it precludes the inversion of a sparse matrix. The integral can be removed by dividing by $y^2$ and differentiating with respect to $y$, which yields:
\begin{multline}
    \frac{-2}{y^3} \pder[\phi_s]{y} + \frac{1}{y^2} \dpder[\phi_s]{y} - \frac{1}{\xi \gamma} \pder[\phi_s]{z} = \\ (1 - V_1(\hat{p}_t)) \Big(\frac{1}{y^2} \ddder[s^*]{y}- \frac{2}{y^3} \dder[s^*]{y} \Big)  \label{eq_modified_bc}
\end{multline}
where the entire equation is evaluated at $z^* = 1$. This boundary condition leads to a sparse matrix that can be easily incorporated into a numerical solver.




Before continuing with the numerical solutions, we formulate the SST forcing in the troposphere. We observe from Eq. \ref{eq_steady_vBC} that:
\begin{equation}
    s^* = \int y u_1 \: dy \label{eq_sstar_forcing}
\end{equation}
such that we can specify the baroclinic wind response to obtain a suitable $s^*$ anomaly. Here, we specify:
\begin{equation}
    u_1(y) = -\exp\big( -4 (y - 2)^2 \big) - \exp\big( -4 (y + 2)^2 \big)
\end{equation}
which is akin to subtropical jets symmetric about the equator. Note, the meridional baroclinic wind is:
\begin{equation}
    v_1 = \frac{F}{y} u_0 + \frac{D_t + F}{y^2} \dder[s^*]{y}
\end{equation}
Numerical evaluation of $v_1$ requires that the meridional derivative of $s^*$ go to zero faster than $y^2$ in the limit of $y \rightarrow 0$, otherwise $v_1$ will become unstable for small values of $y$ on the numerical grid. However, the stratospheric solution does not depend on $v_1$, so this constraint merely ensures a smoothly varying tropospheric circulation. Thus, $u_1(y)$ is chosen to satisfy this constraint. We proceed by numerically solving the stratospheric system (Eq. \ref{eq_phi_strat}) with the modified boundary condition shown in Eq. \ref{eq_modified_bc}, as well as the $s^*$ forcing shown in Eq. \ref{eq_sstar_forcing}. See the appendix for more details on the numerical solver.

To set the non-dimensional parameters of the model, we use Earth-like parameters of $N^2~=~6~\times~10^{-4}$~s$^{-2}$, $H = 16$~km, $H_{s,t}~=~H_{s,s}~=~8$~km, $\beta~=~2.3~\times~10^{-11}$~$s^{-1}~m^{-1}$, $L_y~=~1200$ km (such that $y = 1$ represents approximately ten degrees of latitude), $C_d~=~10^{-3}$, $|\textbf{V}|~=~3$~m~s$^{-1}$. Furthermore, we choose $T_b~=~303~\:~\text{K}$, a surface pressure of 1000-hPa, and a tropopause pressure of 100-hPa. The vertical temperature profile in the troposphere follows a pseudoadiabatic lapse rate (neglecting changes to heat capacity, see Eq. 4.7.5 of \citet{emanuel1994atmospheric}), such that $[\overline{T}]~\approx 264.5~\:~\text{K}$ and $\overline{T}(p_t)~\approx~176.1~\:~\text{K}$. With these values, $V_1(p_t)~\approx~-2.3$. 

\begin{figure*}
  \noindent\includegraphics[width=39pc]{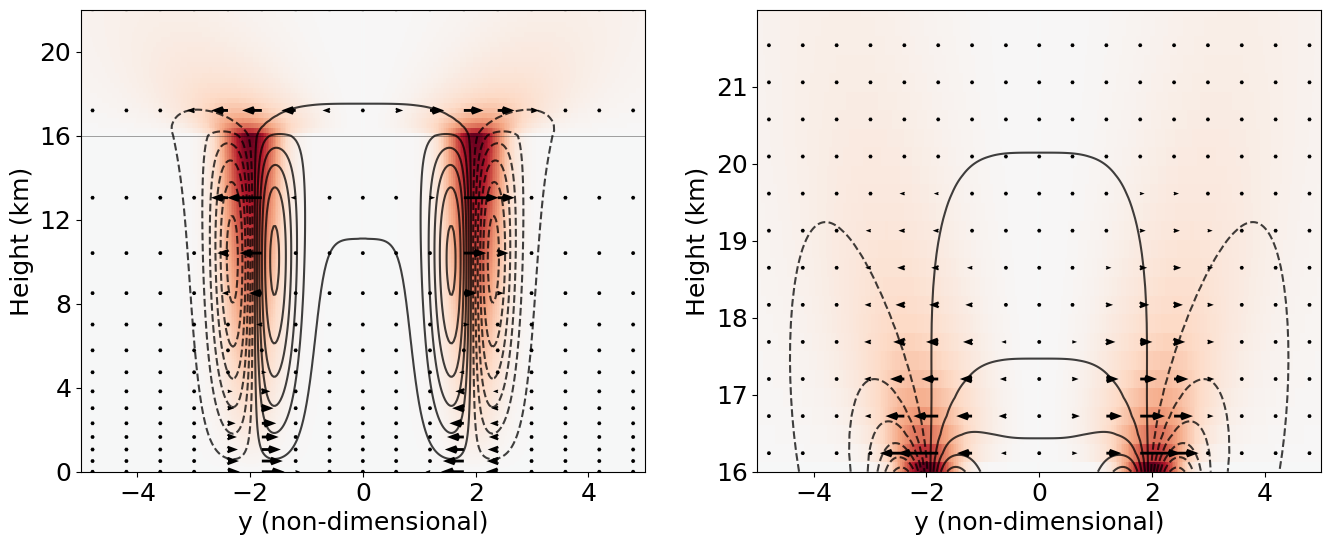}\\
  \caption{(Left): The zonally symmetric response to a SST ($s^*$) forcing shown in Eq. (\ref{eq_sstar_forcing}). Zonal winds are shown in colors (red for westerlies), contours show vertical motion ($w$), where 5 contours indicates a difference of one order of magnitude. Arrows show the meridional motion. The tropopause is shown by the thin gray line. ``Earth-like" parameters of $\xi = 150$, $\gamma = 30$ are used. (Right): Same as left, but zoomed in on the stratosphere.}
  \label{fig_zonally_symmetric_response}
\end{figure*}

Since $\alpha_{\text{rad}}$ and $\hat{D}_t$ play critical roles in the stratospheric response to an imposed tropopause geopotential anomaly, we will explore the the non-dimensional space of $\xi$ and $\gamma$. Still, it is helpful to to note the estimates of the general order of magnitudes of these quantities in the real stratosphere. \citet{hitchcock2010approximation} estimated the radiative relaxation time scale to be approximately 25 days in the lower tropical stratosphere. The magnitude of the Eliassen Palm (EP) flux divergence is around O(1)~m~s$^{-1}$~$\text{day}^{-1}$ in the subtropics, but decays rapidly as one moves equatorward into the deep tropics \citep{randel2008dynamical}. For a perturbation zonal wind speed of O(10) m/s, this corresponds to a Rayleigh damping rate of around 10 days$^{-1}$ and slower.

For now, we restrict the analysis to ``Earth-like" parameters, with $\hat{\alpha}_{\text{rad}} = 25$~days$^{-1}$, and $\hat{D}_s~=~\hat{D}_t~=~25$~days$^{-1}$. This choice leads to $\xi \approx 150$ and $\gamma \approx 30$. Thus, $\xi \gamma$ is large, and the tropopause geopotential can be approximated as simply a multiple of $s^*$. Figure \ref{fig_zonally_symmetric_response} shows the zonally symmetric, linear response to the prescribed, equatorially symmetric SST forcing. We observe a meridionally shallow, thermally direct overturning circulation in the troposphere, associated with sub-tropical jets at $|y| = 2$ that decay exponentially with height into the stratosphere. The tropopause geopotential is elevated in the tropical region ($|y| < 2$) (not shown). Associated with this elevated tropopause geopotential is a weak, meridionally shallow, thermally indirect overturning circulation in the stratosphere, with upwelling around an order of magnitude smaller than peak upwelling in the troposphere. Note that the tropospheric thermally direct overturning circulation in this model is not meant to realistically mimic the Hadley circulation, since linear models do not capture the dynamics of the Hadley circulation \citep{held1980nonlinear}. Rather, its purpose in this model is to understand how tropopause geopotential anomalies associated with tropospheric circulations influence the stratospheric circulation.

\begin{figure*}
  \noindent\includegraphics[width=39pc]{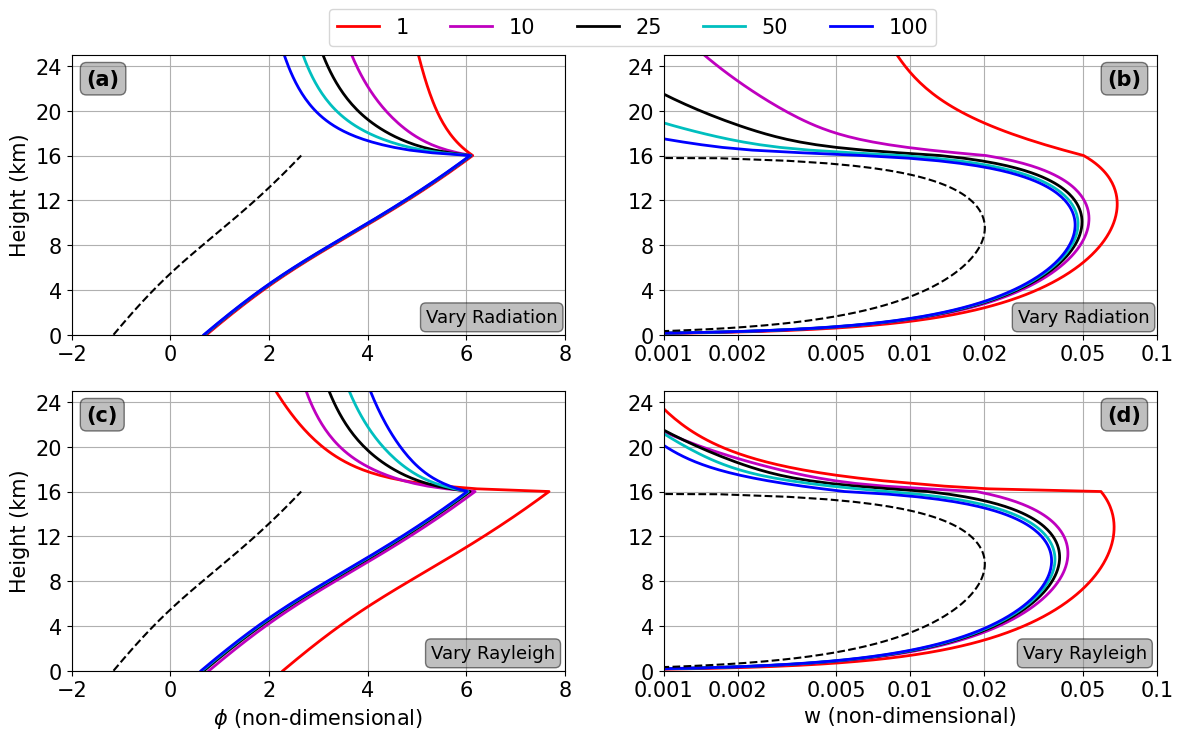} \\
  \caption{\textbf{(a)} Vertical profiles of non-dimensional geopotential and \textbf{(b)} vertical velocity, at y = 1.5, for varying values of radiative relaxation, at a fixed Rayleigh damping (wave drag) of $~25$~days$^{-1}$. Dashed lines show the geopotential and vertical velocity associated with a pure baroclinic mode (normalized so that the peak vertical velocity is $0.02$). \textbf{(c)}, \textbf{(d)} are the same as (a), (b), respectively, except for varying values of stratospheric Rayleigh damping, at a fixed radiative relaxation rate of $25$~days$^{-1}$. Tropopause is defined at 16-km, and tropospheric Rayleigh damping is fixed at $25~$~days$^{-1}$.}
  \label{fig_w_phi_T_profiles}
\end{figure*}

What is the sensitivity of the stratospheric circulation to $\hat{\alpha}_{\text{rad}}$? Figure \ref{fig_w_phi_T_profiles}a,b shows the vertical profile of anomalous geopotential and vertical velocity, for varying values of $\hat{\alpha}_{\text{rad}}$. In all the solutions presented here, the tropospheric wave drag is fixed. We first observe that for all the solutions, the geopotential anomaly maximizes at the tropopause, and there is a significant barotropic geopotential component associated with all of the solutions. These positive geopotential anomalies decay as one moves upwards into the stratosphere, but the rate at which they decay is determined by the aforementioned parameters. When $\hat{\alpha}_{\text{rad}} = 1$~day$^{-1}$, we observe a slow decay of the tropopause geopotential as one moves upwards into the stratosphere, and large upwelling values in the lower stratosphere. In contrast, when radiation is very slow ($\hat{\alpha}_{\text{rad}} = 100$~day$^{-1}$), there is almost no penetration of the tropospheric vertical velocity into the stratosphere. This is associated with a tropospheric vertical velocity profile that is nearly entirely composed of the first baroclinic mode. As expected, radiative damping plays a large role in the communication of the tropopause forcing into the stratosphere.  

The stratospheric response to a steady tropopause geopotential anomaly also shows a strong dependence to $\hat{D}_s$. This is not surprising, given the criticality of wave-drag in the zonally-symmetric solutions. Figure \ref{fig_w_phi_T_profiles}c,d shows the solutions with varying $\hat{D}_s$ and a fixed radiative damping time scale. The behavior of the coupled solutions are qualitatively similar to that inferred from the isolated stratosphere solutions, in that faster wave-drag time scales increase the decay of the tropopause geopotential into the stratosphere. In addition, faster wave-drag time scales are associated with increased upwelling in the lower stratosphere, though the differences across the parameters shown are smaller in magnitude than that when varying the radiative damping time scale. This result could be a result of the simple relaxational form of wave-drag used in this study, which does not capture detailed aspects of wave-forcing \citep{ming2016response}. Regardless, the numerical solutions confirm the mathematical analysis, in that both radiative damping and wave-drag can modulate the stratospheric response to tropospheric forcing. Note, in a similar linear system, PE99 found solutions to a stratosphere perturbed through tropospheric thermal forcing that showed stratospheric upwelling nearly comparable in magnitude to that of the troposphere, which was deemed as unrealistic. In PE99, the radiative relaxation time scale was 10~days$^{-1}$ and the relaxational wave-drag time scale was 500~days$^{-1}$, which corresponds to small $\xi$, and large penetration of the tropospheric circulation into the stratosphere.

\begin{figure*}[t]
  \noindent\includegraphics[width=39pc]{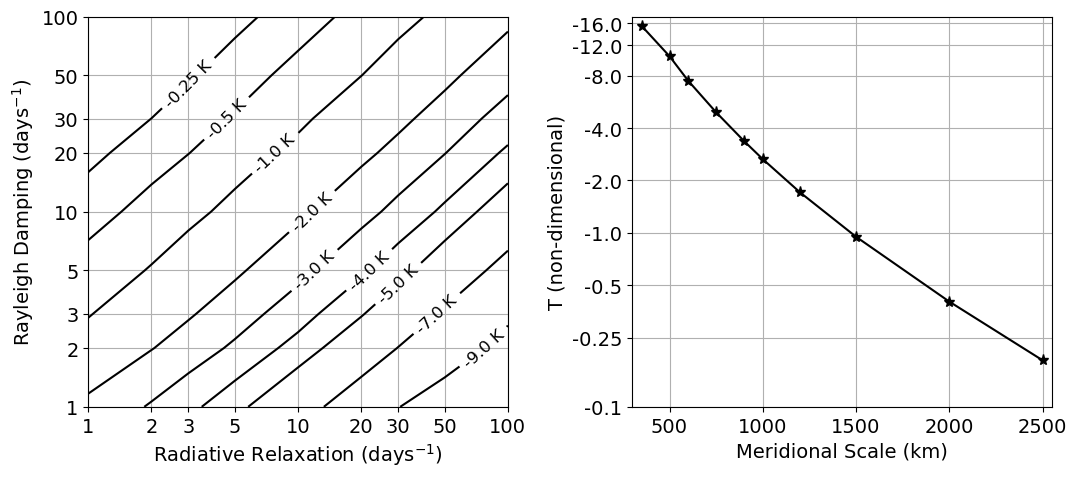} \\
  \caption{(Left): Temperature anomaly right above the tropopause, per degree of warming in the boundary layer, as a function of the radiative relaxation and Rayleigh damping (wave-drag) time scales. Rayleigh damping time scale is fixed in the troposphere and varied in the stratosphere. (Right): Temperature anomaly right above the tropopause, per degree of warming in the boundary layer, as a function of the meridional length scale, $L_y$ (km), for fixed Rayleigh damping and radiative relaxation. Ordinate axis is logarithmic.}
  \label{fig_trop_T_sensitivity}
\end{figure*}

The vertical shape of the geopotential profiles above the tropopause also allows for an estimate of the magnitude of the tropopause temperature cold anomaly as a function of tropospheric heating. Figure \ref{fig_trop_T_sensitivity}, left, shows the temperature anomaly right above the tropopause, per degree of warming in the boundary layer, as a function of the radiative damping and Rayleigh damping time scales. In general, the longer the radiative damping time scales, the larger the temperature anomaly (as pointed out by \citet{randel2002time}). In addition, there is also a strong dependence of the tropopause temperature anomaly on the Rayleigh damping time scale: the faster the damping, the larger the magnitude of the temperature anomaly. It is clear that both the magnitudes of the Rayleigh damping (wave-drag) and radiative damping play significant roles in modulating the temperature anomaly above the tropopause. 

Interestingly, for ``Earth-like" estimates of the time scale of Rayleigh damping and radiative relaxation (O(10) days$^{-1}$), the temperature anomalies just above the tropopause are around 2-3 times the magnitude of the boundary layer anomalies, slightly larger than what is observed in convecting regions of the tropical atmosphere (see Fig. 5a in \citet{holloway2007convective}). This theory thus provides a scaling argument for the degree of tropopause cooling that is expected per degree of boundary layer warming. Note that the derivative of the geopotential is discontinuous across the tropopause in this model, since we assume a instantaneous transition between quasi-equilibrium thermodynamics in the troposphere, and dry, passive dynamics in the stratosphere.

These theoretical results provide a potential explanation for the observed correlation between tropical-averaged SST anomalies and tropical stratospheric upwelling \citep{lin2015tropical}, as well as the anti-correlation between SST and tropopause temperature \citep{holloway2007convective, fu2006enhanced}. First, an SST anomaly is communicated throughout the depth of the troposphere through moist convection. Indeed, observations have found strong positive correlations between the tropopause geopotential anomaly and the boundary layer temperature anomaly \citep{holloway2007convective}. The tropopause geopotential anomaly is initially associated with cold temperature anomalies just above the tropopause. The strength of radiative relaxation then determines the time scale at which the geopotential anomaly rises in the stratosphere through diabatic heating. In the zonally-symmetric case, the presence of wave-drag, through conservation of angular momentum, disrupts this process and induces a meridional overturning circulation that mediates the vertical scale at which the geopotential anomaly can rise in the stratosphere. 


Our work shows that, at least in the zonally symmetric case, the ratio between the strength of radiative relaxation and that of Rayleigh damping are significant factors in determining the response of the stratosphere to an SST anomaly. However, there are a number of other quantities unveiled through the non-dimensionalization that are also important. Surface friction, for instance, factors into $\gamma$. In general, increasing the magnitude of $F$ does little to change the behavior of the stratospheric response to tropospheric forcing when $\xi$ is large, since $F$ only enters in $\gamma$ and $\xi \gamma$ is what matters for the tropopause boundary condition. The tropospheric \& stratospheric stratification, as well as the shape and length scale of the SST (or tropopause) perturbation ($L_y$), also factor into the non-dimensional parameters that control the vertical decay scale of tropopause geopotential anomalies. The horizontal scale of the SST anomaly can also be quite important, due to the dependence of $S$ on $L_y^{-4}$. Figure \ref{fig_trop_T_sensitivity}, right shows the dependence of the temperature anomaly above the tropopause on $L_y$. There is an approximately logarithmic scaling of the temperature anomaly with the meridional length scale of the tropopause anomaly, at least across the range of $L_y$ in the experiments. Correspondingly, the geopotential response in the stratosphere is muted for small $L_y$ (not shown). Thus, large horizontal scale tropospheric heating anomalies have a larger penetrative depth into the stratosphere, but are also associated with smaller (in magnitude) temperature anomalies at the tropopause.

\begin{figure*}[t]
  \noindent\includegraphics[width=39pc]{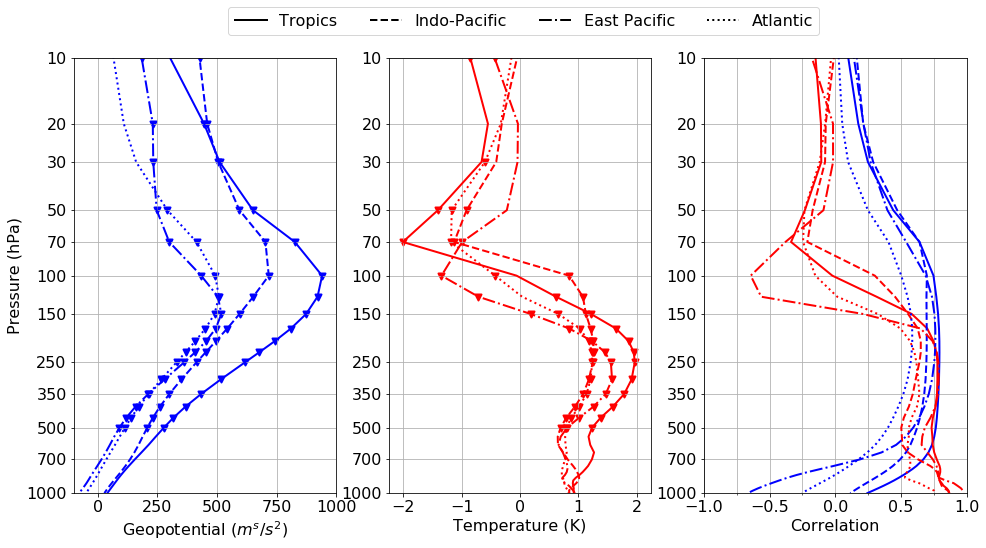} \\
  \caption{(Left) Linear coefficient of geopotential at varying levels, regressed onto regionally-averaged SST anomaly. Above 500-hPa, significant correlations at the 1\% level (two-sided) are denoted by upside-down triangles. (Middle) Same as the left panel but for temperature. (Right) Vertical dependence of the correlation coefficients for (blue) geopotential and (red) temperature. The regions are (solid) the entire tropics [$20^\circ$S - $20^\circ$N], (dashed) the Indo-Pacific region [$40^\circ$E-$120^\circ$E], (dot-dashed) the East Pacific region [$180^\circ$E-$260^\circ$E], and (dotted) the Atlantic region [$80^\circ$E-$0^\circ$]. Vertical level is scaled as the logarithm of pressure.}
  \label{fig_column_regression}
\end{figure*}

\section{Tropopause forcing in reanalysis data \label{sec_reanalyis}}
In this section, we evaluate the proposed theory using the ERA5 re-analysis \citep{hersbach2019era5_single, hersbach2019era5_pressure}. We use monthly fields of SST, geopotential, and temperature, over the years 1979-2022. The Quasi-Biennial Oscillation (QBO) is regressed out of the geopotential and temperature fields, by using the 50-hPa zonal wind averaged over the tropics. In particular, we will analyze correlations between metrics of tropospheric warming and stratospheric cooling, on the global scale and the local scale.

\begin{figure*}
  \noindent\includegraphics[width=39pc]{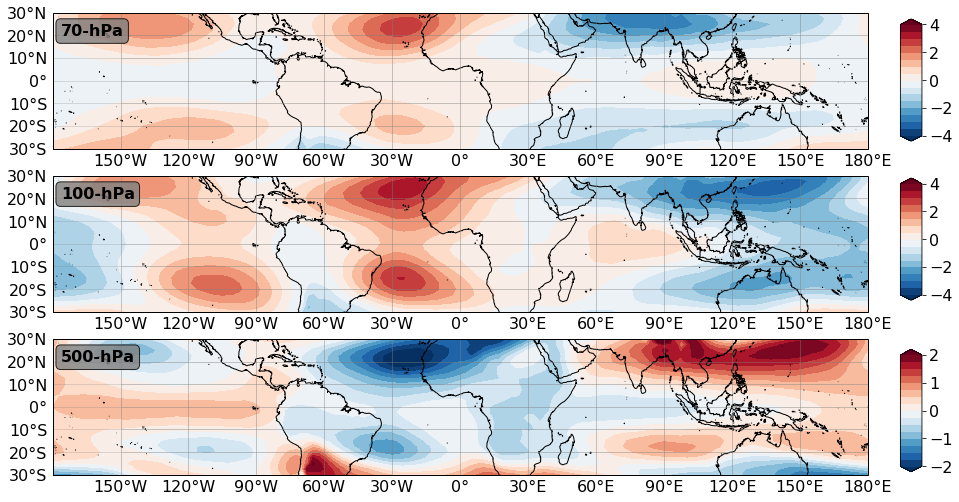} \\
  \caption{Annual-average temperature anomaly at (top) 70-hPa, (middle) 100-hPa, (bottom) 500-hPa. Note the strong anti-correlation in troposphere and lower stratospheric temperature. Anomalies are calculated by subtracting the climatological monthly zonal mean, and averaging across the entire year.}
  \label{fig_temp_correlation}
\end{figure*}

To begin, we regress the anomalous tropical-averaged geopotential, at different vertical levels, onto the tropical-averaged SST anomaly. Anomalies are generated by subtracting the linear trend in each field, as well the seasonal cycle. Figure \ref{fig_column_regression}, solid lines, shows the coefficients of the linear regressions of geopotential and temperature onto SST. We first observe an approximate moist-adiabatic structure in the tropical tropospheric geopotential, consistent with quasi-equilibrium and the findings of previous studies \citep{holloway2007convective}. We also see a large, significant correlation ($r \approx 0.75$) between tropical-averaged SST and the corresponding 100-hPa geopotential. The magnitude of the geopotential anomaly maximizes at 100-hPa, which is interpreted as an approximate tropopause level, since below this level there is warming, and above this level there is cooling (this is not exact, since the cold-point tropopause could occur above this level). Note the similarity to the geopotential profile shown in Figure \ref{fig_w_phi_T_profiles}, which also maximizes around the climatological tropopause. This is indicative of a tropopause geopotential anomaly that is induced by an SST anomaly. The coefficient magnitudes and correlations decay with increasing height in the stratosphere, but are still statistically significant and non-negligible even at 20-hPa.  Note, for a pure baroclinic mode anomaly, the surface geopotential would be anti-correlated with the upper troposphere anomaly (and the SST). Thus, when the surface geopotential is positively correlated with the upper tropospheric anomaly, there is significant barotropic component to the geopotential profile. We indeed observe that the tropical-averaged surface geopotential is positively correlated with both SST and the upper tropospheric geopotential, highlighting the role of the barotropic mode and the troposphere's communication with the stratosphere.

The temperature structure of the tropical troposphere is also approximately moist-adiabatic, as also shown in \citet{holloway2007convective}. Figure \ref{fig_column_regression} also shows that the tropics-averaged 70-hPa temperature is modestly but significantly anti-correlated ($r \approx -0.34)$ with surface temperature. We also observe temperature anomalies at 70-hPa (lower stratosphere) to be approximately two times larger in magnitude than that of the surface, which is in agreement with the estimates shown in Figure \ref{fig_trop_T_sensitivity}. This is not exactly equivalent with the quantity derived in the left portion of Figure \ref{fig_trop_T_sensitivity}, since the poor vertical resolution of the ERA5 reanalysis prohibits an exact determination of the tropopause height.

The same relationships are also observed on regional scales (the Indo-Pacific, East Pacific, and the Atlantic), as shown in Figure \ref{fig_column_regression}. The geopotential anomalies maximize at 100-hPa in the Indo-Pacific, at 125-hPa in the Atlantic, and at 150-hPa in the East Pacific. Thus, the level at which the geopotential anomaly maximizes is influenced by the mean SST of the region (the East Pacific has the coldest climatological SSTs, while the Indo-Pacific has the warmest). In addition, the cold anomaly associated with SST warming maximizes above the level of maximum geopotential. The regional scale geopotential anomalies persist upwards to around 50-hPa, though the correlations drop significantly in magnitude, and the statistical significance ceases around 50-hPa. This means that regional and local scale variations in the lower stratospheric geopotential (50-hPa and 70-hPa) are strongly influenced by the tropopause geopotential in the same region. 

Of course, this analysis is not definitive proof that there is a quasi-balanced response of the stratosphere to tropopause forcing. After all, if stratospheric temperature is modulated by tropical heating through changes to wave-drag \citep{garcia2008acceleration, calvo2010dynamical, lin2015tropical}, then one would also expect the geopotential to decay with height in the stratosphere, as is shown in Figure \ref{fig_column_regression}. Perhaps what would serve as stronger evidence for the processes described in this study is if the spatial signature of tropospheric warming is retained in that of stratospheric cooling.

In the tropics, the surface temperature need not always be connected to tropospheric warming, especially if the boundary layer moist static energy is lower than the saturation moist static energy of the free troposphere. This is possible since temperature gradients in the tropical atmosphere are weak, owing to the smallness of the Coriolis force, such that convecting regions more effectively modulate the free tropospheric moist static energy \citep{sobel2000modeling}. Furthermore, the aforementioned analyses lose information on spatial correlation, since the anomalies are averaged over regions. In order to emphasize spatial variability, we compute monthly anomalies by subtracting the climatological monthly zonal mean from the climatological monthly mean, and then average these across all 12 months (the analysis can also be performed on each month, as will be discussed later).

\begin{figure*}
  \noindent\includegraphics[width=39pc]{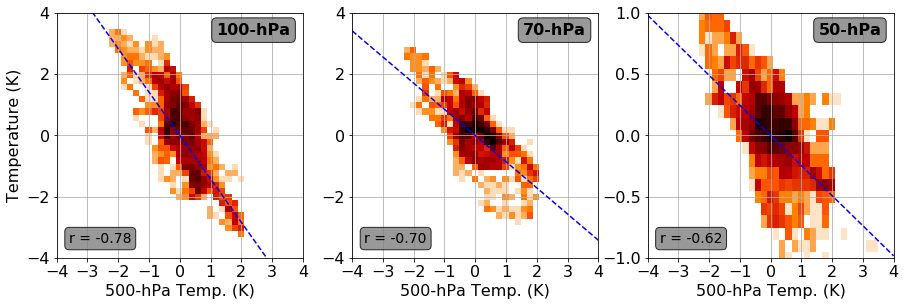} \\
  \caption{Grid-point 2D density histograms between the 500-hPa anomalous temperature and the (left) 100-hPa, (middle) 70-hPa and (right) 50-hPa anomalous temperature. Color scale is logarithmic. Linear regressions are plotted as the dashed blue lines, with correlation coefficients shown on the lower left of each panel.}
  \label{fig_temp_regressions}
\end{figure*}

Figure \ref{fig_temp_correlation} shows a map of the 500-hPa temperature anomaly, a proxy for tropospheric heating, and the anomalous temperature at 100- and 70-hPa in the lower stratosphere. It is evident that 500-hPa temperature is an excellent predictor of both the 100-hPa and 70-hPa temperature anomaly. Spatial variability in the tropospheric temperature anomaly is remarkably retained in the spatial variability of the stratospheric temperature. Furthermore, the lower-stratospheric temperature anomalies can be rather large (upwards to around 4 degrees in magnitude at 100-hPa and 70-hPa), though the total area encompassed by these large anomalies is small. There is also some qualitative evidence from the maps in Figure \ref{fig_temp_correlation} that suggests that the magnitude of the lower stratospheric temperature anomalies is dependent on the horizontal scale of the tropospheric anomaly. For instance, from 60$^\circ$W to 20$^\circ$E in the Northern Hemisphere, there is a large-scale tropospheric cold anomaly of peak magnitude around 2 degrees. The associated temperature anomaly at 100-hPa is around 4 degrees. There is also a large-scale tropospheric warm anomaly of peak magnitude around 2 degrees in the Asian region (60$^\circ$E to 180$^\circ$E), with 100-hPa temperature anomalies of around -4 degrees. In contrast, smaller scale tropospheric anomalies [(150$^\circ$W to 90$^\circ$W, 10$^\circ$S to 30$^\circ$S), (45$^\circ$W to 15$^\circ$W, 10$^\circ$S to 25$^\circ$S)] with comparatively weaker peak temperature anomalies are associated with 100-hPa temperature anomalies that are of similar magnitude to the 100-hPa temperature anomalies of the stronger, large scale anomalies. This is in agreement with the proposed theory. In addition, at 70-hPa, the most prominent temperature anomalies are those associated with the large-scale tropospheric anomalies (i.e. over the Northeast African and Asian regions). This is also in agreement with the theory, in that the vertical depth of the tropopause anomalies increases with the horizontal scale of the tropospheric anomaly. Of course, the analysis here is mostly qualitative, and could be marred by the poor vertical resolution of the reanalysis. More substantial analysis is required to further quantify the scale dependence of the lower stratospheric temperature anomalies, and will be pursued in future work.

The remarkable correlation between tropospheric heating and stratospheric cooling can be further quantified by regressions of 500-hPa temperature against lower-stratospheric temperature, among all grid points shown in Figure \ref{fig_temp_correlation}. Figure \ref{fig_temp_regressions} shows 2-D density histograms between the 500-hPa temperature and the 100-, 70-, and 50-hPa temperature, as well as the linear regressions. Per degree of warming at 500-hPa, the cooling response is around 1.5 degrees at 100-hPa $(r = -0.78)$, 1 degree at 70-hPa $(r = -0.7)$, and 0.25 degrees at 50-hPa $(r = -0.62)$. The correlations are all significant, and generally decrease in strength as one moves up further in the stratosphere. While the monthly anomalies shown in Figure \ref{fig_temp_correlation} are averaged across the whole year, there is significant seasonal variability in the pattern of 500-hPa tropospheric temperature (not shown). The analysis can repeated by separating into each month, and the results and interpretation remained unchanged: 500-hPa temperature is strongly anti-correlated with lower stratospheric temperature. It is important to note that these correlations do not suggest that there are correlations on significantly smaller horizontal scales; as suggested by Figure \ref{fig_temp_correlation}, the correlations merely reflect the large-scale structure of the temperature anomalies. Regardless, these simple analyses provide strong evidence that there is a quasi-balanced response of the stratosphere to tropospheric thermal forcing.

\section{Summary and discussion \label{sec_conclusion}}
In this work, we present theoretical evidence for how tropopause geopotential anomalies, generated through tropospheric thermal (SST) forcing, can modulate upwelling in the stratosphere. Using a conceptual model based on the linearized QGPV equations, we show that tropospheric thermal forcing can induce a tropopause geopotential anomaly, which subsequently elicits a quasi-balanced response in the stratosphere. The tropopause anomalies initially have vertically shallow structures scaled by the Rossby penetration depth (i.e. the fast adjustment of the stratosphere). Afterwards, radiative relaxation in the stratosphere acts to increase the vertical penetration of these anomalies. In the steady-state limit, where radiative equilibrium is again satisfied, the stratospheric PV becomes barotropic, though it takes on the order of years to be achieved. The solutions are akin to those of \citet{haynes1991downward}, who found that the stratosphere becomes barotropic above the level of forcing (in this case, the tropopause).

We then formulate a zonally symmetric troposphere-stratosphere linear $\beta$-plane model, which couples a convecting troposphere to a dry and passive stratosphere. We show that the stratospheric response to tropospheric forcing is controlled by two non-dimensional parameters: (1) $\xi$, a dynamical aspect ratio \citep{ming2016response}, and (2) $\gamma$, a ratio between the stratospheric drag and tropospheric friction. In the limit that radiation is much stronger than wave drag, the stratospheric response to a tropopause forcing asymptotically becomes barotropic, while in the opposite limit, the vertical length scale of the tropopause forcing becomes extremely small. We find that the stratospheric response to zonally-symmetric tropospheric forcing is largely dependent on the radiative relaxation rate, the Rayleigh damping time scale of wave-drag, and the horizontal scale. Our analyses show that the tropopause temperature anomaly is also modulated by all of these quantities.

We also use reanalysis data to show that tropical and regionally averaged lower-stratospheric temperatures are modestly and negatively correlated with SSTs in the same areas. In general, the temperature anomalies per degree of warming in the boundary layer are approximately equivalent to the corresponding theoretical predictions, at least when using ``Earth-like" estimates of the time scale of wave-drag and radiative relaxation. Furthermore, we show that the spatial variability in lower-stratospheric temperature anomalies is strongly correlated with the spatial variability in 500-hPa tropospheric temperatures. Significant correlations are seen upwards to 50-hPa, which suggests that there is a quasi-balanced response of the stratospheric to tropospheric forcing. This provides a scale-dependent theory for the oft-observed anti-correlation between tropospheric warming and stratospheric cooling \citep{johnson1982thermodynamic, gettelman2002distribution, fu2006enhanced, holloway2007convective, kim2012tropical, virts2014observations, kim2018convectively}.

The widely accepted theory of tropical stratospheric upwelling is that it is mechanically driven by sub-tropical wave-drag \citep{haynes1987evolution, plumb1999brewer}. There is ample evidence from numerical modeling suggesting that wave-dissipation is a dominant mechanism that modulates mean and interannual upwelling in both the lower stratosphere and TTL \citep[among many others]{boehm2003implications, norton2006tropical, calvo2010dynamical, ryu2010effect, gerber2012stratospheric, ortland2014residual, kim2016spectrum, jucker2017untangling}. Of course, it is theoretically impossible to have flow across angular momentum contours without some momentum source. We emphasize that in no way does this work attempt to disprove the role sub-tropical wave drag has in modulating tropical stratospheric upwelling. In this model, even though wave-drag acts as a Rayleigh damping, as in the linear system described in PE99, it is an important modulator of the upwelling response.

As shown in this study, the vertical penetration of the geopotential anomaly (and the rate at which the stratospheric circulation crosses angular momentum surfaces) is strongly a function of the wave drag. If the wave-drag is a function of 
the zonal mean state, which could vary in time in part due to wave-forcing \citep{cohen2013compensation, ming2016response}, then the vertical penetration of the tropopause anomaly (and thus, its subsequent effect on upwelling) would also vary in time. In this view, stratospheric wave-drag is, as countless studies have shown, a significant modulator of tropical upwelling. However, wave drag alone may not suffice to explain certain features of the behavior of the lower stratosphere, the foremost of which is the inverse correlation between SST and lower stratospheric temperature anomalies, in both the zonal and meridional directions.

Our work, like PE99, investigates how tropospheric thermal forcing can modulate stratospheric upwelling. In addition to mechanical and thermal forcing, this suggests a \textbf{third} way in which the stratosphere can be forced -- through the tropopause via tropospheric thermal forcing. In fact, the theoretical analysis shown in PE99 finds that in the tropics, ``the existence of a thermally driven circulation and the breakdown of downward control go together" (if one accepts that what they define as viscosity is representative of large-scale drag). However, their calculation of the linear response to tropospheric thermal forcing exhibited large and unrealistic vertical penetration of the tropospheric circulation into the stratosphere. This work shows that this is likely a result of their assumptions of the strength of radiative relaxation ($\alpha_{\text{rad}} = 10$ days$^{-1}$) and viscosity ($\hat{D} = 500$ days$^{-1}$). With $S = O(10^2)$, this is equivalent to $\xi \approx 3$. In this regime, our theory predicts extensive penetration of the tropospheric circulation into the stratosphere, as in Figure \ref{fig_strat_zonally_symmetric_response} and \ref{fig_w_phi_T_profiles}. 

In general, it is difficult to infer causality from diagnostic relations. For example, in Transformed-Eulerian Mean equations (derived, for instance, in \citet{andrews1987middle}), it is not clear how much of the wave-drag is an external forcing, as opposed to a response to a circulation that has a different forcing. Of course, variations in wave-drag that are independent of those of the circulation support the idea that waves can force the circulation. This aspect of the stratosphere has been well studied. But what if wave-drag acted purely as a response to the circulation? (Note that these ideas are at opposite ends of the spectrum with regards to the extent waves drive the circulation)? Then, at least in our framework, the causality becomes very clear -- SST forces the stratosphere by imposing a tropopause geopotential anomaly. Of course, one could take the wave-drag term ($-D_s u_s$) and use it to diagnose the associated upwelling response. However, that does not imply that waves are the forcing mechanism of the circulation.

There are a few pieces of observational evidence that could be interpreted to be in favor of the proposed theory. As stated earlier, the spatial variability of lower-stratospheric temperature is strongly correlated with that of the troposphere. In addition, there is a strong observed anti-correlation of temperature trends in the troposphere and the lower stratosphere \citep{fu2006enhanced}. These long-term trends are highly correlated on a grid-point by grid-point basis, suggesting that the zonal and meridional structure of tropospheric warming may be important to that of stratospheric cooling. In contrast, wave-drag, in its classical arguments, can only explain departures of temperature from the zonal-mean \citep{andrews1987middle}. This is by no means a small feat, since the annual cycle in tropical-averaged temperature near the tropopause is around 8K, around a factor of two larger than the peak temperature anomalies shown in Figure \ref{fig_temp_correlation} \citep{chae2007annual}.

However, the quasi-balanced response of the stratosphere to tropopause forcing could serve as a potential explanation for a few outstanding issues. For instance, it can explain why there is peak tropical upwelling on the summer-side equator \citet{rosenlof1995seasonal}. It could also help to explain the observed connection between boundary layer temperature anomalies and lower stratospheric temperature anomalies, as well as the high correlations between tropical SST and the upwelling strength of the shallow BDC branch, which is observed on all time scales \citep{lin2015tropical, abalos2021brewer}. Numerical modeling suggests that strengthening of the sub-tropical jets changes the upward propagation of waves \citep{garcia2008acceleration, calvo2010dynamical, shepherd2011robust}, ultimately strengthening the wave-driven stratospheric upwelling, although the exact specifics seem to vary from model to model \citep{calvo2010dynamical, simpson2011dynamics}. In the zonally symmetric coupled troposphere-stratosphere theory analyzed in this work, an equatorial SST anomaly is not only associated with strengthening of the sub-tropical jets (which no doubt could change the sub-tropical distribution of wave-drag in the real-world), but also a strengthening of the tropopause geopotential. As such, the theory proposed in this work does not have to be mutually exclusive with those based on wave-drag. 

Besides the inclusion of a relaxational wave-drag (shown to be a poor assumption), our work stays silent on how the momentum budget must change in order to balance changes in the meridional circulation \citep{ming2016response}. However, there would undoubtedly be a large scale wave response to steady tropospheric heating \citep{gill1980some}. Thus, disentangling the effects of heating from the ensuing wave-response is quite complicated, as the two occur in concert. While other studies have analyzed the wave-response to tropospheric heating \citep{ortland2014residual, jucker2017untangling} (as well as its subsequent effects on the stratospheric circulation), we have instead focused on the \textbf{steady} response to tropospheric heating. In general, however, when tropical tropospheric heating is used to generate a wave response, it is difficult to separate the tropopause forcing mechanism described in this study from wave driving. For instance, \citet{jucker2017untangling} used idealized GCM simulations to show that the inclusion of a tropical warm pool significantly changed the annual-mean temperature of the tropical tropopause (and more importantly, more so than mid-latitude land-sea contrast and orographic forcing). However, the imposition of a warm pool will both intensify the tropopause anti-cyclone over the region, and trigger a large-scale wave response. According to the analysis shown in this study, the increased tropopause geopotential will act to cool the tropopause and induce more upwelling (as would increased wave-drag from the large-scale wave response). Separately, \citet{ortland2014residual} forced equatorial waves by prescribing time-varying latent heating anomalies in a primitive equation model, and found that stationary waves and weakly westward propagating waves are most responsible for driving residual-mean upwelling in the TTL. Again, tropospheric heating will induce a tropopause geopotential anomaly, such that the steady tropospheric forcing is not separated from the wave response. Regardless, both of the modeling results in \citet{ortland2014residual} and \citet{jucker2017untangling} show that at least in numerical models, the seasonal cycle in upwelling in the tropical tropopause layer cannot be explained by tropospheric thermal forcing. 

It is only fair for these conclusions to be discussed alongside the assumptions posited in this model. In this model, we assume that there is an instantaneous transition between tropospheric, quasi-equilibrium dynamics, and passive, dry stratospheric dynamics. In reality, the presence of the TTL could dampen the upwards influence of tropospheric forcing. The assumption of a moist adiabatic lapse rate all the way to the tropopause is one that is has mixed observational evidence, which suggests that the free tropospheric temperature anomalies, per degree of warming in the boundary layer, approximately follow a moist adiabat up to around 200-hPa, after which temperature anomalies transition to being out of phase with lower tropospheric temperature anomalies [see Figure \ref{fig_column_regression} and \citet{holloway2007convective}] (though some of this may be owing to time averaging with a vertically moving tropopause). While the proposed theory can predict the magnitude of the tropopause temperature anomalies with respect to boundary layer warming, it does not include a transition layer. The presence of a transition layer could, in theory, dampen the vertical penetration of thermal forcing in the troposphere. This will be the subject of future research.

Finally, we also assume a fixed tropopause height that interfaces the two regimes, as in PE99. This makes the analysis mathematically tractable. Indeed, one would expect tropospheric temperature to affect tropopause height \citep{held1982height, lin2017changes}. The relaxation of both of these assumptions will be the subject of future research, but requires a theory for how moist convection interacts with the transition layer. More research is necessary to understand the role of convection in modulating the behavior of the transition layer.

The analysis carried out in section \ref{sec_reanalyis} uses the ERA5 reanalysis dataset, which has very coarse vertical resolution near the tropopause. Since tropopause anomalies can decay in the vertical very quickly, especially for anomalies with small horizontal scale, further insight into the processes outlined in this study would be impaired by the poor vertical resolution of the reanalysis. This could be mitigated by the use of GPS radio-occultation (RO) measurements, provided by the COSMIC mission \citep{anthes2008cosmic}. The high vertical resolution of GPS-RO measurements could be leveraged in future work. Furthermore, while we focused on large-scale tropospheric anomalies in this work, there are also numerous mesoscale convective systems, usually with anticyclones at their tops, that might also be able to contribute to tracer transport into the stratosphere. Higher resolution observational data, such as that provided by GPS RO measurements, could also be useful to evaluate this possibility.

%

%

\newpage
\acknowledgments The author thanks Adam Sobel and Peter Hitchcock for comments and suggestions on earlier versions of this work. The authors also thank two anonymous reviewers and Peter Haynes for their helpful suggestions, which greatly improved the manuscript. In particular, the authors are grateful for Peter Haynes's suggestions on the formulation of the coupled boundary condition. J. Lin gratefully acknowledges the support of the National Science Foundation through the NSF-AGS Postdoctoral Fellowship, under award number AGS-PRF-2201441.


%
%
\datastatement The monthly-mean ERA5 data for sea-surface temperature is available at \url{https://cds.climate.copernicus.eu/cdsapp#!/dataset/reanalysis-era5-single-levels-monthly-means} via DOI: 10.24381/cds.f17050d7 \citet{hersbach2019era5_single}. The monthly-averaged ERA5 data for temperature and geopotential are available at \url{https://cds.climate.copernicus.eu/cdsapp#!/dataset/reanalysis-era5-pressure-levels-monthly-means} via DOI: 10.24381/cds.6860a573 \citet{hersbach2019era5_pressure}.  All code to generate the data from the theoretical models are available at \url{https://github.com/linjonathan/steady_coupled_trop_strat}.


%






%



\appendix
\appendixtitle{Details on Solutions}
\subsection{Solutions to Conceptual Model in Section \ref{sec_qgpv}}
The general solution to the homogeneous version of Eq. \ref{eq_qgpv_linear} ($q(z) = 0$) is:
\begin{align}
    G(z) = A \exp(m_+ z) + B \exp(m_- z)
    \label{eq_greens_homog}
\end{align}
where $m_\pm = \frac{1 \pm \sqrt{1 + 4 (k^2 + l^2)}}{2}$. Note, since $k^2 + l^2 > 0$, $m_+ > 0$ and $m_- < 0$ for all $k > 0$ and $l > 0$. We next define the Green's function, which satisfies
\begin{equation}
    L G(z, \lambda) = \delta(z - \lambda)
    \label{eq_LGreens}
\end{equation}
and is
\begin{equation}
G(z, \lambda) = \begin{cases}
      A \exp(m_+ z) + B \exp(m_- z), & \text{for } 0 < z < \lambda \\
      C \exp(m_+ z) + D \exp(m_- z), & \text{for } \lambda < z < z_{\text{top}} \\
\end{cases}
\end{equation}
where $z_{\text{top}}$ is assumed to be the top of the domain. The lower boundary condition requires that:
\begin{equation}
    A + B = \phi_T \label{eq_boundary1}
\end{equation}
and the upper boundary condition requires that:
\begin{equation}
    C m_+ \exp(m_+ z_{\text{top}}) + D m_- \exp(m_- z_{\text{top}}) = 0
    \label{eq_boundary2}
\end{equation}
Note that we choose to explicitly include $z_{\text{top}}$ in Eq. \ref{eq_boundary2}, since numerically evaluating the Green's functions requires $z_{\text{top}} < \infty$. Continuity of $G$ across $\lambda$  requires:
\begin{align}
    &A \exp(m_+ \lambda) + B \exp(m_- \lambda) = C \exp(m_+ \lambda) + D \exp(m_- \lambda) \label{eq_cont1} \\
    &\lim_{\epsilon \rightarrow 0} \pder[G]{z} \bigg|_{z = \lambda - \epsilon}^{z = \lambda +\epsilon} - \lim_{\epsilon \rightarrow 0} G \Big|_{z = \lambda - \epsilon}^{z = \lambda +\epsilon} = 1 \label{eq_cont2}
\end{align}
Eqs. \ref{eq_boundary1}-\ref{eq_cont2} are solved to obtain:
\begin{align}
    A = \frac{\phi_T - \frac{1}{m_d} ( \exp(-m_- \lambda) - \frac{m_+}{m_-} \exp(-m_+ \lambda + m_d z_{\text{top}}))}{1 - \frac{m_+}{m_-} \exp(m_d z_{\text{top}})} 
\end{align}
where
\begin{equation}
    m_d = m_+ - m_- = \sqrt{1 + 4 (k^2 + l^2)} > 0
\end{equation}
$B$, $C$, and $D$ are then obtained using Eqs. \ref{eq_boundary1}, \ref{eq_boundary2}, and \ref{eq_cont1}.

The Green's function can be convoluted with the source term ($q$) to obtain the geopotential:
\begin{equation}
    \phi(z) = \int_{0}^{\infty} G(z, \lambda) q(\lambda) d\lambda
\end{equation}
\subsection{Numerical Solver for Coupled System}
Here, we elaborate on the numerical solver of the coupled troposphere-stratosphere system (Eq. \ref{eq_phi_strat}, \ref{eq_modified_bc}), given forcing in $s^*$. We approximate the meridional and vertical derivatives with second-order and sixth-order central finite differences, respectively. Since our specified $s^*$ forcing is equatorially symmetric, we only have to discretize $y$ from equator to pole, and impose a Neumann boundary condition at the equator. However, $y$ appears in the denominator in both Eq. \ref{eq_phi_strat} and \ref{eq_modified_bc}). We circumvent this issue by numerically evaluating the equator at $\epsilon = 10^{-5}$ (three orders of magnitude smaller than the meridional grid spacing). $y$ is evenly discretized from $y_{\text{max}}$ to $\epsilon$, where $y_{\text{max}} = -10$. $z$ is evenly discretized from the tropopause ($z^* = 1$) to the domain top, $z^*_{\text{top}} = 7$. The boundary conditions are:
\begin{align}
    \phi(y = y_{\text{max}}, z^*) &= 0 \\
    \pder[\phi]{y} (y = \epsilon, z^*) &= 0 \\
    \pder[\phi]{z} (y, z^* = z^*_{\text{top}}) &= 0
\end{align}
as well as the aforementioned Eq. \ref{eq_modified_bc} on the boundary $z^* = 1$. The solutions are ensured to solve the original linear system of equations, as well as the boundary conditions, within numerical error. Finally, we use the \textit{findiff} Python package to solve the system numerically \citep{findiff}.

\bibliographystyle{ametsocV6}
\bibliography{references}

\end{document}